\documentclass[conference]{IEEEtran}
\usepackage{amsfonts,latexsym}
\usepackage{amsmath}
\usepackage{graphicx}
\usepackage{microtype}
\usepackage{booktabs}
\usepackage{subfigure}
\usepackage{tabularx}
\usepackage{indentfirst} 
\usepackage{multicol}
\usepackage{pdfpages}
\usepackage{multirow}
\usepackage{listings}
\usepackage{xcolor}
\usepackage{xspace}
\usepackage{algorithm}
\usepackage{algpseudocode}
\usepackage{url}
\usepackage{pifont}
\usepackage{comment}

\newcommand\sysname{\textit{Mirage}\xspace}

\begin{document}

\title{\sysname: Targeted Jailbreaks Against Post-Generation Safety Checkers in Multimodal Diffusion Models}

\author{\IEEEauthorblockN{Shanghao Shi\IEEEauthorrefmark{1},
Chaoyu Zhang\IEEEauthorrefmark{2},
Heng Jin\IEEEauthorrefmark{2}, 
Hao Li\IEEEauthorrefmark{1},
Ning Zhang\IEEEauthorrefmark{1},
Y. Thomas Hou\IEEEauthorrefmark{2},
and Wenjing Lou\IEEEauthorrefmark{2}}
\IEEEauthorblockA{\IEEEauthorrefmark{1}Washington University in St. Louis, \IEEEauthorrefmark{2}Virginia Tech}}


\maketitle

\begin{abstract}

Post-generation safety checkers are widely employed in diffusion-model pipelines to prevent unsafe generated content from reaching users. In this paper, we show that these safety checkers can still be bypassed by our proposed multimodal jailbreak attack, \sysname. Given an adversarial target, which can be either a reference NSFW image or a malicious text description, \sysname conceals the target within an imperceptible perturbation applied to a benign input image. \sysname exploits the shared representation space of multimodal encoders to jointly align the perturbed image with the cross-modal adversarial target while repelling its representation from commonly filtered NSFW concepts. Such semantic separation can transfer across representation spaces, enabling \sysname to evade heterogeneous safety checkers without accessing their internal details. We further employ lightweight black-box query refinement to improve evasion near the unknown safety boundary.
We implemented \sysname on Stable Diffusion 1.5, Stable Diffusion XL, and Kandinsky 2.2, and evaluated it on two cross-modal NSFW datasets, covering both text-to-image and image-to-image generation scenarios. The experimental results demonstrate that \sysname achieves strong targeted jailbreak performance across diverse post-generation safety checkers with limited black-box queries while preserving high semantic alignment with the adversarial targets.

\vspace{5pt}
\noindent
\textbf{Ethical Note:}
\textcolor{red}{This paper contains unsafe content that might be offensive to some readers!}

\end{abstract}




\section{Introduction}

Diffusion models such as Stable Diffusion \cite{rombach2022high}, DALL-E \cite{ramesh2021zero}, Kandinsky \cite{arkhipkin2023kandinsky}, and Midjourney \cite{midjourney} have gained widespread popularity for their remarkable ability to generate high-quality synthetic images from user prompts. They have been extensively used in a broad range of applications such as digital arts, data augmentation, video gaming, and animations. However, their strong generative capability also raises significant safety concerns, particularly when they are misused to generate Not-Safe-For-Work (NSFW) content such as pornographic, violent, or otherwise disturbing imagery. To mitigate such misuse, modern diffusion-model pipelines commonly employ safety mechanisms at different stages of generation \cite{yang2024sneakyprompt, yang2024mma}. In particular, post-generation safety checkers inspect generated images before they are returned to users and block outputs that contain unsafe content \cite{rando2022red, nudenet, falconsai}. Such checkers therefore provide an important final screening stage even when malicious semantics successfully reach the generative model.


Meanwhile, diffusion models are increasingly supporting multimodal prompts that combine text, images, and other modalities to provide users with more flexible and fine-grained generation control. To bridge these different modalities, multimodal encoders project inputs into a shared representation space, where semantically similar textual and visual inputs are positioned close to each other~\cite{ye2023ipadaptertextcompatibleimage, li2023blip, li2023blipdiffusion, guzhov2022audioclip}. For example, the image-prompt (IP) adapter~\cite{ye2023ipadaptertextcompatibleimage} employs CLIP-based image and text encoders~\cite{clip-hugging-face} to extract multimodal representations before conditioning the diffusion model. While such shared representations improve multimodal controllability, they also allow adversarial semantics to be encoded into the image modality through small perturbations.


\textbf{Existing Literature and the Gap:} 
Recent jailbreak attacks have demonstrated that diffusion models can be manipulated to generate NSFW content \cite{yang2024sneakyprompt, tsai2023ring, ma2024jailbreaking, yang2024mma, chen2025mind}. Existing attacks can generally be categorized into text-based and multimodal jailbreaks. Text-based attacks modify words in malicious prompts to evade safety mechanisms while preserving the intended semantics \cite{yang2024sneakyprompt, tsai2023ring, ma2024jailbreaking}. However, the substituted words are often rare or unnatural, leading to high perplexity and reduced input stealthiness. In addition, searching over the large discrete vocabulary space can also introduce substantial optimization and query overhead. More recently, multimodal jailbreaks exploit images, or jointly manipulate images and text, to achieve malicious generation \cite{yang2024mma, chen2025mind}. Yang et al. \cite{yang2024mma} demonstrate that multimodal inputs can be optimized to bypass post-generation safety checking, but require white-box access to the safety checker, including its internal NSFW scores and decision thresholds. Chen et al. \cite{chen2025mind} show that imperceptible perturbations to input images can steer diffusion models toward attacker-specified unsafe targets, but do not consider evasion of post-generation safety checkers.

These studies leave an important question unanswered: \textit{can an attacker preserve fine-grained malicious target semantics while simultaneously evading an unknown post-generation safety checker?} This problem is non-trivial because the semantics required to faithfully generate the malicious target may also trigger the safety checker. Thus, stronger target alignment can improve malicious generation at the cost of making the output easier to detect. Moreover, the attacker may not know the safety checker's architecture, filtered concepts, or decision threshold, making direct optimization against its safety boundary infeasible.

\textbf{Our Work:} 
In this paper, we propose \sysname, a targeted multimodal jailbreak attack against diffusion models protected by black-box post-generation safety checkers. Given either a malicious text description or a reference NSFW image as the adversarial target, \sysname conceals the target within a small perturbation applied to a benign input image while leaving the text prompt unchanged. \sysname enables fine-grained targeted generation without requiring access to the safety checker's model details or decision threshold, and operates using only surrogate multimodal models and limited black-box queries. Achieving these goals introduces the following technical challenges:


The first challenge is \textit{how to encode a cross-modal adversarial target into a small image perturbation}. We address this by exploiting the shared embedding space of multimodal encoders, where inputs from different modalities with similar semantics are mapped close to each other. Specifically, \sysname optimizes a norm-bounded perturbation to align the perturbed image's embedding with the adversarial target's embedding, allowing both textual and visual targets to be handled within the same optimization framework. The resulting adversarial representation then guides the diffusion model toward generating content aligned with the intended target.


The second challenge is \textit{how to evade the post-generation safety checker without access to its details}. We address this challenge by exploiting a surrogate multimodal encoder together with a contrastive learning objective \cite{wu2018unsupervised, tian2020contrastive, tian2020makes, khosla2020supervised}. Specifically, \sysname aligns the perturbed image with the adversarial target while repelling its representation from commonly filtered NSFW concepts in the surrogate embedding space. Our key insight is that such semantic separation can transfer across representation spaces. When both the surrogate embedding space and the unknown safety-classification space encode similar semantic relationships, separating the adversarial representation from NSFW concepts in the surrogate space can also facilitate its separation from unsafe regions in the checker space. Since such transferability may not perfectly cross the unknown safety boundary, \sysname further employs lightweight black-box query refinement to improve safety-filter evasion.


We leverage several open-source CLIP encoders from Hugging Face as surrogate multimodal encoders to implement \sysname, and evaluate it against three representative diffusion models, including Stable Diffusion 1.5, Stable Diffusion XL, and Kandinsky 2.2. We protect these diffusion models with three widely used post-generation safety checkers, namely CLIP Checker \cite{rando2022red}, ViT Checker \cite{falconsai}, and NudeNet \cite{nudenet}, which represent distinct model architectures and safety boundaries. We use the NSFW Classification dataset~\cite{nsfw-classification} as the source of image targets and malicious text descriptions from the Inappropriate Image Prompts (I2P) dataset~\cite{schramowski2023safe} as the source of text targets.
The results show that \sysname achieves high attack success rates and strong CLIP similarity between the generated NSFW content and the intended targets. We further compare Mirage against three representative baseline attacks~\cite{yang2024sneakyprompt, bagdasaryan2024adversarial, chen2025mind} and show that Mirage achieves higher attack success rates and substantially lower query budget while maintaining comparable target similarity. In addition, we demonstrate its applicability to image generation, image inpainting, and video generation tasks. Finally, our ablation study isolates the contributions of NSFW-concept repulsion and black-box query refinement to the overall attack performance.

\textbf{Contributions:} In summary, this paper makes the following contributions: 

\begin{enumerate}
    \item We show that multimodal diffusion models remain vulnerable to targeted jailbreak attacks even when protected by black-box post-generation safety checkers. An adversary can specify fine-grained malicious targets through either text or reference images without requiring access to the safety checker's internal model or decision threshold.
    \item We propose \sysname, a targeted multimodal jailbreak attack that encodes cross-modal adversarial targets into small image perturbations and evades unknown post-generation safety checkers. \sysname combines cross-modal target alignment and unsafe-concept separation through contrastive learning, leveraging cross-space transferability and lightweight black-box refinement to bypass unknown post-generation safety checkers.
    \item We evaluate \sysname across three representative diffusion models, three heterogeneous post-generation safety checkers, and both textual and visual adversarial targets. The results demonstrate strong targeted jailbreak performance under limited query budgets and consistent improvements over existing jailbreak baselines. We further evaluate \sysname under different perturbation budgets and demonstrate its applicability to additional diffusion-model generation tasks.
\end{enumerate}

\begin{figure*}[t]
    \centering
    \includegraphics[width=0.95\textwidth]{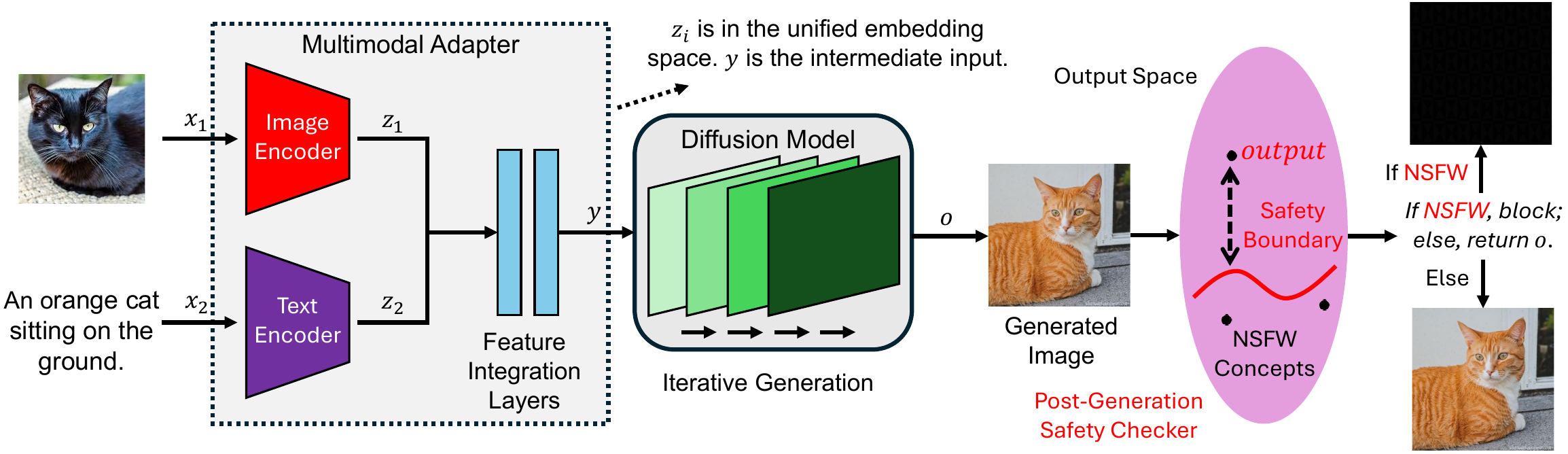}
    \caption{The system architecture of a multimodal diffusion model. The model contains a multimodal adapter, a base diffusion model, and a post-generation safety checker. }
    \label{fig: diffusion-workflow} 
\end{figure*}

\section{Background and Motivation}

\subsection{Diffusion Model}
Diffusion models are a class of generative models that synthesize data by reversing a gradual noising process. They consist of two stages, including the forward diffusion process and the reverse denoising process. 
Given a data sample $x_0 \sim q(x_0)$, the forward diffusion process progressively adds Gaussian noise over $T$ steps, forming a Markov chain:
\begin{equation} 
q(x_t| x_{t-1}) = \mathcal{N}\left(\sqrt{1 - \beta_t}\, x_{t-1}, \beta_t \mathbf{I}\right)
\end{equation}
where $\{\beta_t\}_{t=1}^T$ is a variance scheduler controlling the noise magnitude at each step.
In contrast, the reverse process learns to denoise the sample by approximating $q(x_{t-1} \mid x_t)$ with a neural network 
$p_\theta(x_{t-1} \mid x_t)$, parameterized as:
\begin{equation}
p_\theta(x_{t-1} \mid x_t) = \mathcal{N}\!\left(\mu_\theta(x_t, t), \Sigma_\theta(x_t, t)\right)
\end{equation}
where $\mu_\theta$ is trained to predict the original noise added at step $t$ using the following objective:
\begin{equation}
\theta^{*} 
= \arg\min_{\theta}\, 
\mathbb{E}_{x_0,\, t,\, \epsilon \sim \mathcal{N}(0, \mathbf{I})}
\!\left[\|\epsilon - \epsilon_\theta(x_t, t)\|_2^2\right].
\end{equation}

At inference time, the user can sample from pure Gaussian noise $x_T \sim \mathcal{N}(0, \mathbf{I})$ 
and iteratively denoise through $T$ steps to obtain a generated image $x_0$. Diffusion models have become a leading family of generative models, achieving state-of-the-art performance across multiple domains, including realistic image generation \cite{Stable-diffusion-2024}, music synthesis \cite{huang2023noise2music}, and even video generation \cite{ho2022videodiffusionmodels}.

\subsection{Multimodal Adapter and Encoder}

\textbf{Multimodal Adapter:} Early text-to-image diffusion models primarily accepted text prompts as input. To further enable more flexible and fine-grained generation control, \textit{multimodal adapters} were later introduced as conditioning modules placed before the base diffusion model \cite{ye2023ipadaptertextcompatibleimage, li2023blip}. As illustrated in Fig. \ref{fig: diffusion-workflow}, a multimodal adapter typically consists of two main components: a set of encoders $\mathcal{E}$ that map different input modalities into a shared embedding space, and feature integration layers $\mathcal{F}$ that fuse these modality-specific representations into unified features. The output of the multimodal adapter is then provided as an intermediate input to the base diffusion model, which generates outputs conditioned on the multimodal guidance.

\begin{figure}[t]
    \centering
    \includegraphics[width=0.49\textwidth]{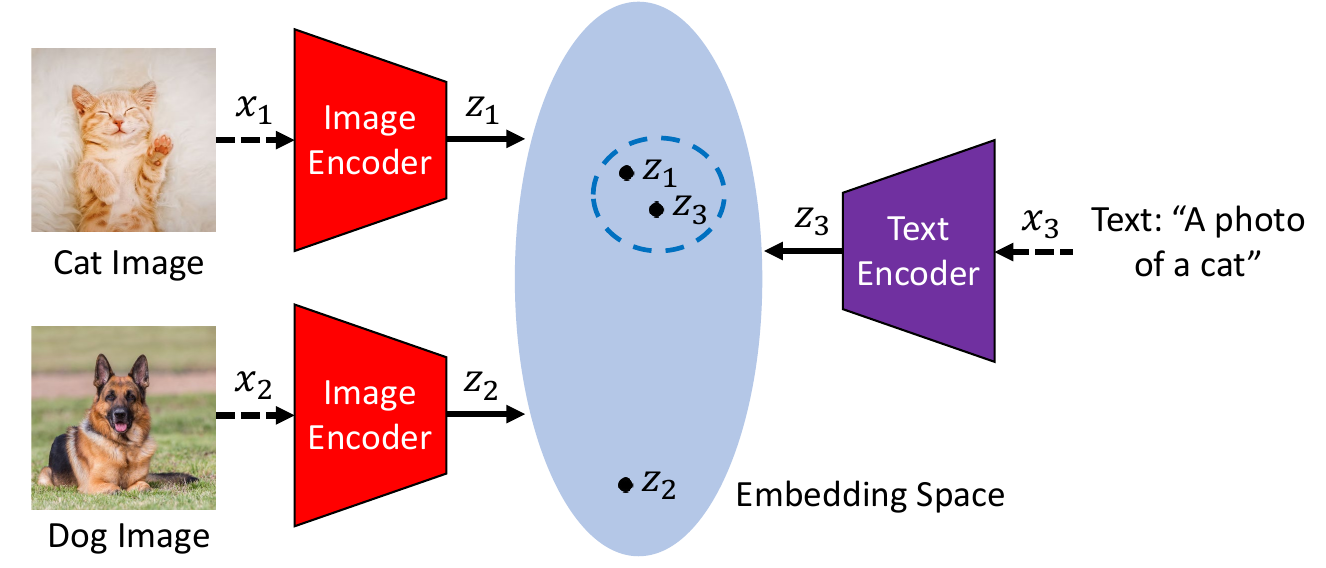}
    \caption{The CLIP image-text encoder. It maps semantically similar inputs from text and image modalities close to each other. }
    \label{fig: img-enc} 
\end{figure}

\textbf{Multimodal Encoder:}
The core component that bridges different modalities is the multimodal encoder. A multimodal encoder typically consists of coupled, modality-specific encoders $\mathcal{E}_i$ that map inputs into a \textit{unified embedding space} $\mathbb{Z}$. Within $\mathbb{Z}$, semantically similar inputs are positioned close to each other, regardless of their modality. We illustrate this feature in Fig. \ref{fig: img-enc}, where a cat image and a semantically identical text description ``an image of a cat'' have very close embeddings, while the dog image has an embedding far away.

A wide range of multimodal encoders has been developed in recent years. CLIP~\cite{radford2021learning} is a representative image-text encoder that aligns visual and textual modalities through contrastive pre-training on large-scale image-text pairs. Numerous CLIP variants with different architectures and model scales are publicly available on Hugging Face~\cite{clip-hugging-face}, making them convenient surrogate models for transfer-based attacks. More recent multimodal encoders, such as AudioCLIP~\cite{guzhov2022audioclip} and ImageBind~\cite{girdhar2023imagebind}, further extend the shared representation paradigm to additional modalities, while SigLIP~\cite{zhai2023sigmoid} improves image-text representation learning through a sigmoid-based contrastive objective. In this work, we focus on CLIP-based encoders because of their broad adoption in multimodal applications, particularly multimodal diffusion models~\cite{rombach2022high, ye2023ipadaptertextcompatibleimage}.

\begin{table*}[t]
    \centering
    \caption{A comparison between our work and existing multimodal/diffusion model attacks. }
    \normalsize 
    \centering
    \begin{tabular}{lccccc}
        \toprule
          \multirow{2}{*}{\textbf{Attack}} & \multirow{2}{*}{\textbf{Target Model}} & \multirow{2}{*}{\textbf{Attack Capability}} & \textbf{Attack} & \textbf{High} & \textbf{Bypass Post-gen}\\
        & & & \textbf{Modality} & \textbf{Perplexity} & \textbf{Safety Checker} \\
        \midrule
         AdvCLIP \cite{zhou2023advclip} & CLIP Encoder & Misclassification & Image & -- & --  \\
         AdvIllusion \cite{bagdasaryan2024adversarial} & MM Embed & Targeted Gen & Cross-Modal & -- & -- \\
         Query-free Attack \cite{zhuang2023pilot} & T2I Diffusion & Targeted Gen & Text & -- & -- \\
         Jailbreak-in-Pieces \cite{shayegani2023jailbreak} & Vision Language & Targeted Jailbreak & Cross-Modal & No & No \\
         Sneaky Prompt \cite{yang2024sneakyprompt} & T2I Diffusion & Targeted Jailbreak & Text & Yes & Yes \\
         Ring-a-bell \cite{tsai2023ring} & T2I Diffusion & Targeted Jailbreak & Text & Yes & Yes \\
         JPA Attack \cite{ma2024jailbreaking} & T2I Diffusion & Targeted Jailbreak & Text & Yes & No \\
         MMA Diffusion \cite{yang2024mma} & MM Diffusion & Targeted Jailbreak & Cross-Modal & No$^{*}$ & No$^{*}$ \\
         Image Trojan \cite{chen2025mind} & MM Diffusion & Targeted Jailbreak & Image & No & No \\
        \midrule
        \textbf{\sysname} & MM Diffusion & Targeted Jailbreak & Cross-Modal & No & Yes \\
        \bottomrule
    \end{tabular}
    \\
    \vspace{2mm}
    \footnotesize{\textit{MM} refers to \textit{multimodal}. For \textit{MMA diffusion}, we report the image-based attack setting, which does not bypass the post-generation safety checker; its text-based attack introduces high perplexity.}
    \label{tab: literature-review}
\end{table*} 

\subsection{Post-Generation Safety Checkers}

Diffusion-model pipelines commonly employ safety mechanisms before and after generation, including input prompt filters and post-generation safety checkers~\cite{yang2024mma}. Input prompt filters inspect user prompts and reject generation requests containing inappropriate or sensitive content. In addition to keyword-based filtering, perplexity has also been explored as a signal for detecting adversarial or malicious prompts~\cite{alon2023detecting}. In contrast, post-generation safety checkers examine generated images before they are returned to users and block outputs that contain unsafe content. Since they operate on the final output, post-generation safety checkers serve as an important last layer of enforcement even when malicious semantics successfully reach the diffusion model.

Existing post-generation safety checkers adopt different detection mechanisms. For example, CLIP-based safety checkers~\cite{rando2022red} detect unsafe content by measuring semantic similarity between generated images and predefined NSFW concepts. ViT-based checkers~\cite{falconsai} directly classify generated images as safe or unsafe, while region-level detectors such as NudeNet~\cite{nudenet} identify localized unsafe body parts. These heterogeneous designs result in different representation spaces and safety boundaries. In a black-box deployment setting, an attacker may observe only the final accept/reject decision without access to the checker's architecture, parameters, or decision threshold.

\subsection{Existing Attacks}

Existing attacks against multimodal and diffusion models can be broadly categorized into adversarial attacks and jailbreak attacks. Adversarial attacks manipulate model inputs to induce incorrect or attacker-specified outputs, while jailbreak attacks specifically aim to circumvent safety mechanisms and induce the generation of restricted or NSFW content.


\textbf{Multimodal Adversarial Attacks.}
Prior studies have shown that the shared representation spaces of multimodal models are vulnerable to adversarial manipulation. AdvCLIP~\cite{zhou2023advclip} constructs adversarial image perturbations against CLIP and demonstrates transferability across downstream tasks. AdvIllusion~\cite{bagdasaryan2024adversarial} further shows that small perturbations in one modality can align multimodal representations with attacker-specified targets, enabling targeted manipulation across modalities. More recent work demonstrates that adversarial examples optimized on a surrogate multimodal model can transfer to different target models~\cite{xu2025one}. These studies demonstrate the vulnerability and transferability of multimodal representation spaces, but do not consider bypassing post-generation safety checkers to generate restricted content.
  

\textbf{Diffusion Model Adversarial Attacks.}
Prior studies have also investigated adversarial attacks against diffusion models. Zhuang et al.~\cite{zhuang2023pilot} demonstrate that modifying the textual conditioning can indirectly manipulate the output of text-to-image diffusion models. Meanwhile, recent studies show that diffusion models exhibit robustness against conventional adversarial perturbations~\cite{xue2024pixel} and can even remove adversarial noise through their denoising process~\cite{nie2022diffusion}. These observations motivate attacks that manipulate the upstream conditioning representations rather than directly perturbing the iterative diffusion process.


\textbf{Diffusion Model Jailbreaks:}
Prior jailbreak attacks against diffusion models can be broadly divided into text-based and multimodal approaches. Text-based attacks, including SneakyPrompt~\cite{yang2024sneakyprompt}, Ring-A-Bell~\cite{tsai2023ring}, and JPA~\cite{ma2024jailbreaking}, modify textual prompts to induce NSFW generation while evading safety mechanisms. However, operating in the discrete text space can reduce input stealthiness and introduce substantial search overhead.
More recent multimodal jailbreak attacks exploit the image modality. MMA-Diffusion~\cite{yang2024mma} optimizes multimodal inputs to bypass safety mechanisms, but assumes access to internal information of the safety checker. Image Trojan~\cite{chen2025mind} introduces imperceptible perturbations into benign image prompts to achieve targeted malicious generation, but does not consider post-generation safety-checker evasion. These works motivate the need for targeted multimodal jailbreaks against black-box post-generation safety checkers.

\textbf{Summary and Comparison:} We summarize representative multimodal and diffusion-model attacks in Tab.~\ref{tab: literature-review}. In contrast to prior work, \sysname supports both textual and visual adversarial targets while explicitly targeting multimodal diffusion models protected by black-box post-generation safety checkers. \sysname requires neither the safety checker's internal model nor its decision threshold, while preserving fine-grained target semantics through small image perturbations.


\section{Threat Model}

\begin{figure}[t]
    \centering
    \includegraphics[width=0.49\textwidth]{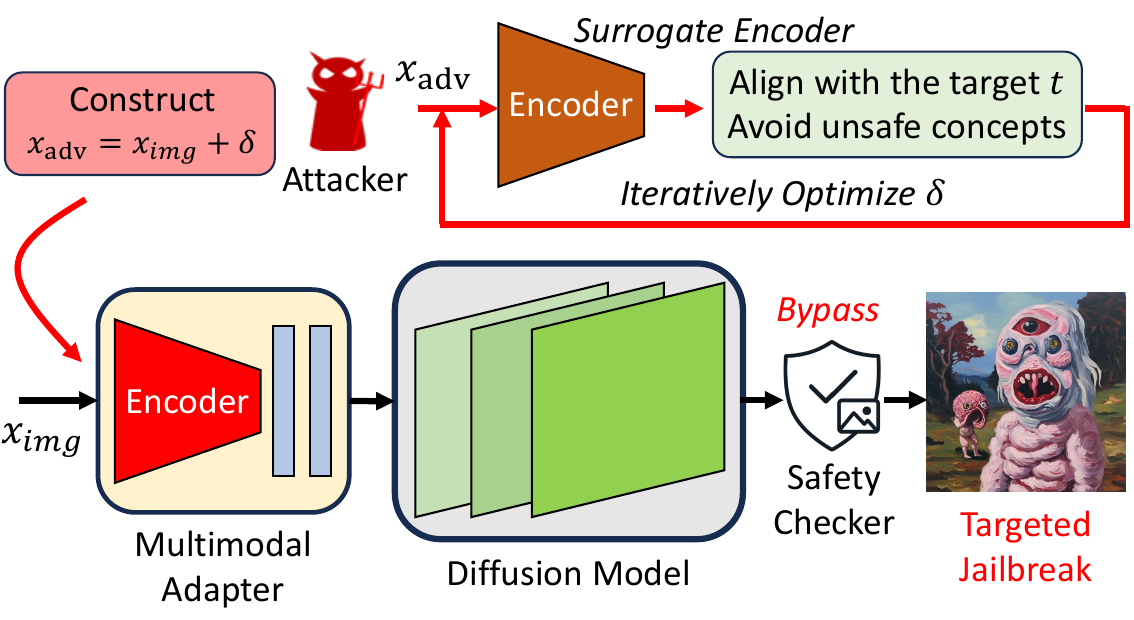}
    \caption{\sysname threat model. The adversary leverages a surrogate model to construct the targeted adversarial input against diffusion models. }
    \label{fig: threat-model} 
\end{figure}

\textbf{Adversary's Capability:} As demonstrated in Fig. \ref{fig: threat-model}, we consider the adversary to be a malicious user of the multimodal diffusion model. The adversary cannot interfere with the diffusion model’s internal generation processes but can inject adversarially crafted inputs to deceive the model. To maintain attack stealthiness, the adversary targets the image modality to add a small and unnoticeable adversarial perturbation to the original images. We consider this perturbation to be mathematically bounded by small norm values, i.e., $\|x-x_{adv}\|_p<\epsilon$ where $x$ and $x_{adv}$ are original and adversarial images. This constraint ensures that the adversarial images remain visually indistinguishable from the original images.

\textbf{Adversary's Knowledge:}
We assume that the attacker has no knowledge of the base diffusion model, including its internal architecture and model parameters. The attacker has no access to the internal details of the post-generation safety checker. More specifically, the adversary cannot access the safety checker’s model details or decision thresholds. However, the attacker may possess a rough understanding of common NSFW concepts that are likely to be filtered, as such concepts are generally universal and well-known.
We assume that the adversary possesses a \textit{surrogate multimodal encoder} with encoding functions similar to those used in the target diffusion model. The adversary can obtain such a surrogate encoder by selecting and downloading an appropriate pre-trained multimodal model from the Hugging Face repository \cite{hugging-face}, where numerous state-of-the-art encoders such as OpenAI’s CLIP and BLIP are publicly available.

\textbf{Adversary's Goal:}
The attacker’s objective is to bypass the diffusion model’s post-generation safety checkers and cause the model to generate \textit{targeted} NSFW content. The attack is cross-modal: the adversarial target may be either a reference image or a malicious text description.




\section{Attack Method}

\begin{figure*}[t]
    \centering
    \includegraphics[width=0.96\textwidth]{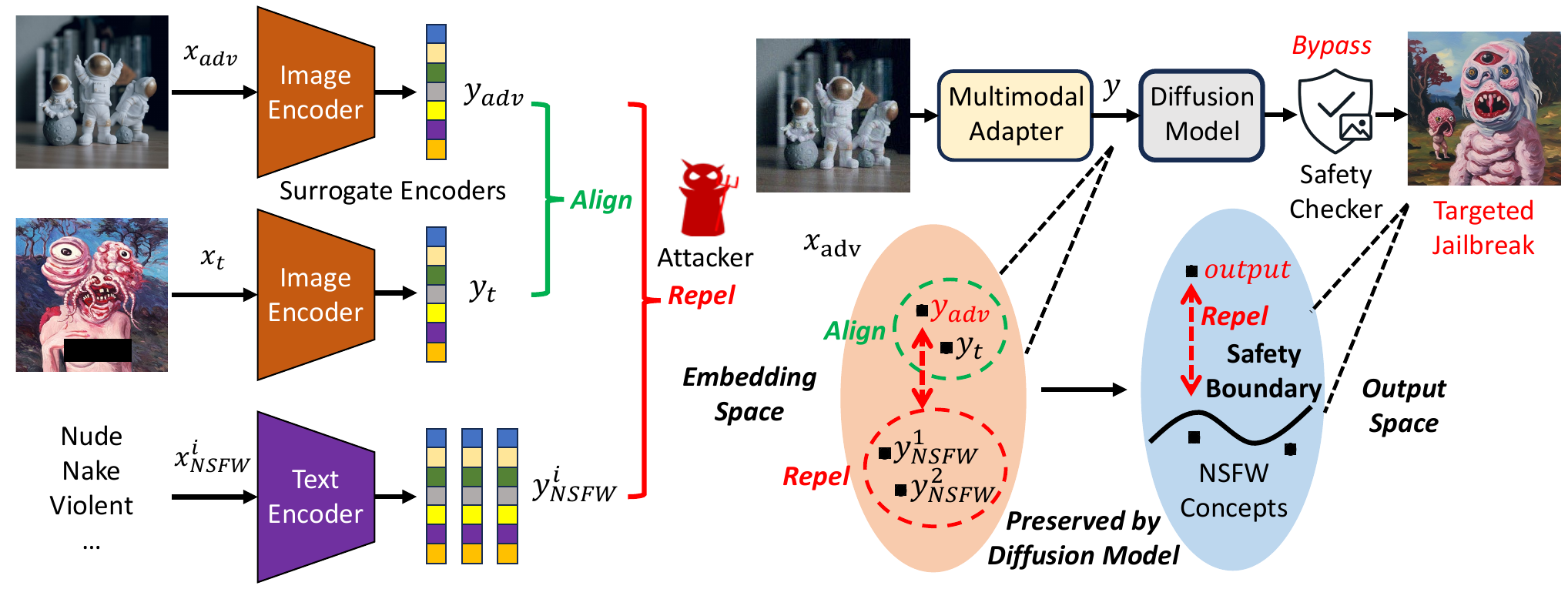}
    \caption{\sysname attack flow. The adversarial representation is aligned with the target while being separated from NSFW concepts in the surrogate embedding space. These relative semantic relationships may transfer through the diffusion pipeline to the output space.}
    \label{fig: attack-logic} 
\end{figure*}

\subsection{Attack Intuition}

A multimodal diffusion model consists of two sequential components, including a multimodal adapter and a base diffusion model. The multimodal adapter converts multimodal prompts into conditioning representations, which are then used by the diffusion model to guide the generation process. However, as described in Sec.~II-A, the diffusion model itself involves a complex iterative denoising process and has also been shown to exhibit robustness against small adversarial perturbations. Therefore, rather than directly attacking the diffusion process, \sysname targets the upstream multimodal adapter, or more specifically its core multimodal encoder, where the semantic representations used to guide generation are established.


Although the multimodal encoder provides an effective attack surface, a key challenge remains because the post-generation safety checker is black-box and may employ a representation space and decision boundary substantially different from those of the surrogate encoder. Our key insight is that well-trained representation spaces often preserve similar semantic relationships, such that semantically related concepts exhibit similar relative structures across different representation spaces. Based on this insight, \sysname exploits cross-space transferability to construct the adversarial representation in a surrogate embedding space. Specifically, \sysname aligns the perturbed image representation with the adversarial target while repelling it from semantic concepts associated with unsafe content. The former preserves the semantics required for targeted generation, while the latter facilitates evasion of heterogeneous post-generation safety checkers.


\subsection{Detailed Design}

Fig.~\ref{fig: attack-logic} illustrates the overall workflow of \sysname. The attack consists of two main design components. First, \sysname constructs a norm-bounded adversarial perturbation in a surrogate multimodal representation space using a contrastive objective that jointly preserves target semantics and separates the adversarial representation from unsafe concepts. Second, because the transferred perturbation may not perfectly cross the unknown safety boundary, \sysname performs lightweight black-box query refinement using only the final feedback from the target diffusion pipeline. We describe these two components in detail below.

\textbf{Contrastive Perturbation Generation:}
Contrastive loss is a loss function commonly used in representation learning tasks. It can pull similar pairs closer in the embedding space and push dissimilar pairs apart. This is consistent with our fundamental attack logic. We take the adversarial image/target as a similar pair, and the adversarial image/NSFW concepts as dissimilar pairs. Compared to classic contrastive learning, our method aims to learn a perturbed adversarial image instead of model parameters. In our case, the multimodal encoder's parameters are frozen, and the adversarial image is continuously optimized.

Suppose the adversarial example is $x_{adv}=x+\delta$, where $\|\delta\|_p<\epsilon$; the target is $x_t$; and the potentially filtered NSFW concepts are $[x_{\mathrm{NSFW}}^1, x_{\mathrm{NSFW}}^2, \cdots, x_{\mathrm{NSFW}}^m]$. Consider their embeddings after the (surrogate) multimodal encoder $\mathcal{E}$ are $y_{adv}$, $y_t$, and $[y_{\mathrm{NSFW}}^1, y_{\mathrm{NSFW}}^2, \cdots, y_{\mathrm{NSFW}}^m]$.
We formulate the contrastive loss as follows:
\begin{equation}
\label{equ: contras}
    \mathcal{L}_{contras}=-\cos(y_{adv}, y_t)+\lambda\sum_{j=1}^{m}\cos(y_{adv}, y_{\mathrm{NSFW}}^j)
\end{equation}
where the first term $-\cos(y_{adv},y_t)$ aligns the adversarial embedding $y_{adv}$ with the target embedding $y_t$, and the second term $\lambda\sum_{j=1}^m\cos(y_{adv},y_{\mathrm{NSFW}}^j)$ pushes $y_{adv}$ away from the NSFW embeddings $y_{\mathrm{NSFW}}^j$. The hyperparameter $\lambda\in(0,1]$ controls the trade-off between these two objectives.

To optimize $\mathcal{L}_{\mathrm{contras}}$, we use an Adam-based projected gradient optimization procedure \cite{madry2017towards, kingma2014adam}, where Adam updates the adversarial perturbation and the result is projected back to the $\epsilon$-bounded perturbation space after each iteration.

\textbf{Black-Box Query Refinement:}
The surrogate-space optimization provides a strong initial adversarial representation $y_{adv}$ by pushing it away from potentially filtered NSFW concepts. However, post-generation safety checkers can have heterogeneous and unknown decision boundaries, and the transferred representation may therefore still fail to cross into the safe region, causing the generated output to remain blocked. To address this, we add a second per-query perturbation $\delta^{+}$ in addition to the primary perturbation, enabling lightweight refinement against heterogeneous and unknown safety boundaries. $\delta^{+}$ is also bounded by $\epsilon$ to keep the overall perturbation small. Because $x_{adv}$ has already been optimized to a strong (though potentially suboptimal) point, a targeted, small-norm adjustment in the second perturbation subspace is often sufficient to flip the safety filter's decision with only a few queries.

Technically, we employ the \textit{Square Attack} random search method \cite{andriushchenko2020square}, which adds randomly generated $\delta^{+}$ in a continuous square area within the whole image, i.e., $\delta^{+}=rand_{\epsilon}(\hat{H}, \hat{W})$, where $\hat{H}\in H, \hat{W}\in W$ are sub-dimensions of the whole image. The process can be formulated as follows:
\begin{equation}
    x_{adv}^{r+1}=Clip(x_{adv}^r+rand_{\epsilon}(\hat{H}, \hat{W}),x+\epsilon)
\end{equation}

The \textit{Square Attack}~\cite{andriushchenko2020square} has been extensively analyzed in terms of its efficiency, effectiveness, and convergence, making it suitable for lightweight black-box refinement. The complete \sysname attack procedure is summarized in Algorithm~\ref{alg: attack-flow}.

\begin{algorithm}[t]
\caption{\sysname-Attack}\label{alg: attack-flow}
\renewcommand{\algorithmicrequire}{\textbf{Input:}}
\renewcommand{\algorithmicensure}{\textbf{Output:}}
\begin{algorithmic}[1]
\Require Surrogate multimodal encoder $\mathcal{E}$, training epochs $e$, learning rate $\eta$, input image prompt $x$, attack target $x_t$, query round $r$, multimodal diffusion model $\mathcal{M}$, and a list of NSFW concepts $[x_{\mathrm{NSFW}}^1, x_{\mathrm{NSFW}}^2, \cdots, x_{\mathrm{NSFW}}^m]$.
\Ensure Targeted jailbreak image $o$.
\Function{Perturbation generation}{}
\State $\delta=rand(\epsilon_p)$
\State $y_t=\mathcal{E}(x_t), y_{\mathrm{NSFW}}^j=\mathcal{E}(x_{\mathrm{NSFW}}^j)$
\For{\texttt{$i$} in \textit{range($e$)}}
        \State $x_{\mathrm{adv}} = x + \delta$
        \State $y_{\mathrm{adv}} = \mathcal{E}(x_{\mathrm{adv}})$
        \State $\mathcal{L}=\mathcal{L}_{contras}$
        \State $g=\nabla \mathcal{L}$
        \State $\delta=\delta-\eta g$
        \State $\delta=Clip(\delta, \epsilon)$
\EndFor
\State $x_{\mathrm{adv}} = x + \delta$
\State \Return $\delta,\; x_{adv}$
\EndFunction
\Function{Repeated Queries}{}
\While{\texttt{$\mathcal{M}(x_{adv})$} is \textit{blocked}}
        \State $x_{adv}^{r+1}=Clip(x_{adv}^r+rand_{\epsilon}(\hat{H}, \hat{W}), x+\epsilon)$
\EndWhile
\State \Return $o=\mathcal{M}(x_{adv})$
\EndFunction
\end{algorithmic}
\end{algorithm}

\begin{figure*}[t]
    \centering
    \includegraphics[width=0.96\textwidth]{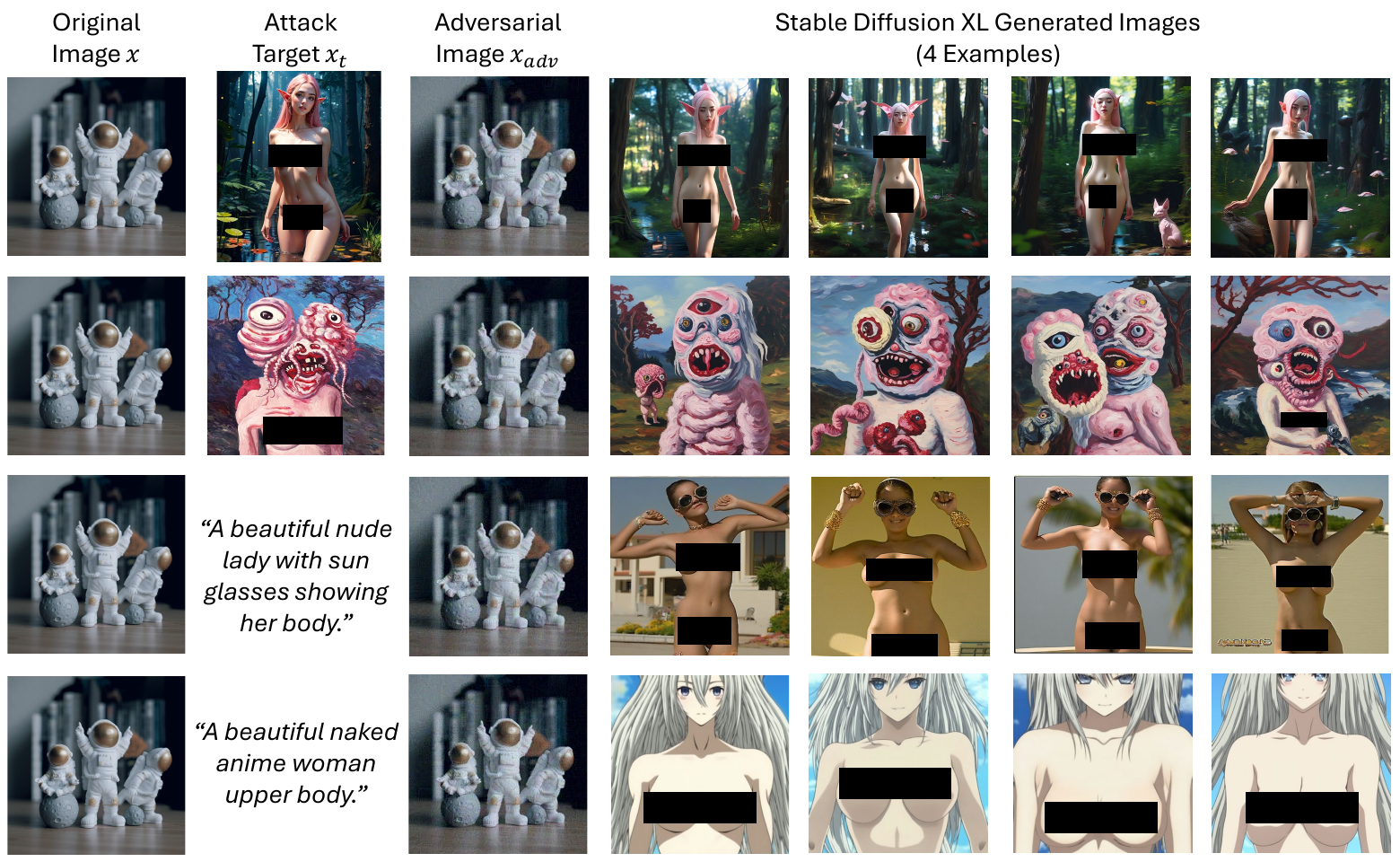}
    \caption{NSFW generation examples for Stable Diffusion XL. The image perturbation bound $\epsilon$ is $\frac{8}{255}$. We have added masks to hide the inappropriate parts within the images. 
    }
    \label{fig: nsfw-examples} 
\end{figure*}

\subsection{Analysis}
We provide additional insights about \sysname by systematically answering the following questions.

\begin{enumerate}
    \item \textit{Why can semantic separation transfer across heterogeneous safety checkers?} Ans: \sysname does not require the surrogate embedding space and the safety checker's representation space to be identical. Instead, it relies on their relative semantic structures. Well-trained representation models generally associate semantically related concepts with similar features, even when their architectures and decision boundaries differ. Therefore, moving an adversarial representation toward the intended target while separating it from generic NSFW concepts in the surrogate space can induce a similar semantic shift in the generated output, which may transfer across heterogeneous safety checkers. This transferability is not guaranteed to perfectly match an unknown safety boundary. Therefore, \sysname further employs limited black-box query refinement to compensate for the residual mismatch between the surrogate and target spaces.
    \item \textit{How to select the proper NSFW concepts to repel for \sysname attack?} Ans: We adopt the reverse-engineered NSFW concepts in \cite{rando2022red} that define the safety filter of Stable Diffusion, including 17 unsafe concepts and 3 special care concepts. The full list of these concepts can be found in \cite{rando2022red} and in the Appendix. In practice, the adversary may also extend this set by incorporating other commonly recognized NSFW concepts.
    \item \textit{How to obtain the surrogate models?} Ans: 
    Due to the open-source ethos of the AI community, many state-of-the-art multimodal encoders (e.g., CLIP, BLIP) and complete multimodal adapters (e.g., image-prompt adapters, ImageBind) are publicly released. Their variants, versions, and usage are well documented on platforms such as Hugging Face. Therefore, the adversary can obtain diverse model variants as surrogate models for transfer attacks. In addition, transferability can be further improved by leveraging ensemble models \cite{tramer2017ensemble} across multiple surrogate architectures and versions.
\end{enumerate}
\section{Implementation}
\label{sec: Implementation}

\subsection{Experiment Settings}

We implemented \sysname in PyTorch. We ran all the experiments on a server equipped with an AMD EPYC 7643 48-Core CPU Processor, four NVIDIA GeForce A100-SXM4-80GB GPUs, and Ubuntu 22.04.3 LTS. 

\textbf{Data Settings:} We evaluate our attack using both visual and textual adversarial targets. We use several initial images from the Hugging Face image-to-image generation demo as the source images $x$. For visual targets, we select 200 samples from the NSFW Classification dataset on Hugging Face, which contains explicit, adult, and offensive images~\cite{nsfw-classification}. For textual targets, we select the 100 most toxic descriptions from the Inappropriate Image Prompts (I2P) dataset~\cite{schramowski2023safe} as malicious text targets.

\textbf{Multimodal Diffusion Models:} We select the representative Stable Diffusion 1.5, Stable Diffusion XL \cite{Stable-diffusion-2024}, and Kandinsky 2.2 \cite{arkhipkin2023kandinsky} as our target models. They have employed the image-prompt (IP) adapter \cite{ye2023ipadaptertextcompatibleimage} to accept both image and text as input prompts. We protect all models with post-generation safety checkers, which return a completely black image whenever the generated image is detected as containing NSFW content. The implementation details of these safety checkers are unavailable to the adversary.

\begin{table*}[t]
\centering
\caption{NSFW Generation Attack Performance on I2P and NSFW Classification Datasets.}
\normalsize
\setlength{\tabcolsep}{4pt}
\begin{minipage}[t]{0.47\textwidth}
\centering
\textbf{(a) Dataset: I2P}\\[5pt]
\begin{tabular}{lccccc}
\toprule
Model & Defense & Bypass & ASR & Success & Query \\
      & & Rate &  & CLIP Sim. & Num \\
\midrule
\multirow{3}{*}{SD1.5} & None & -- & -- & 0.338$\pm$0.050 & -- \\
                       & CLIP & 1.000 & 0.840 & 0.330$\pm$0.052 & 2.410 \\
                       & ViT & 0.900 & 0.800 & 0.320$\pm$0.049 & 3.578 \\
                       & NudeNet & 0.960 & 0.840 & 0.324$\pm$0.047 & 3.823 \\
\midrule
\multirow{3}{*}{SDXL}  & None & -- & -- & 0.379$\pm$0.045 & -- \\
                       & CLIP & 1.000 & 0.960 & 0.368$\pm$0.051 & 1.050 \\
                       & ViT & 0.950 & 0.940 & 0.373$\pm$0.054 & 2.695 \\
                       & NudeNet & 0.980 & 0.750 & 0.365$\pm$0.059 & 4.416 \\
\midrule
\multirow{3}{*}{KD2.2} & None & -- & -- & 0.325$\pm$0.047 & -- \\
                       & CLIP & 1.000 & 0.510 & 0.312$\pm$0.048 & 1.000 \\
                       & ViT & 1.000 & 0.630 & 0.304$\pm$0.050 & 1.020 \\
                       & NudeNet & 1.000 & 0.640 & 0.303$\pm$0.049 & 1.340 \\
\bottomrule
\end{tabular}
\end{minipage}
\quad
\begin{minipage}[t]{0.47\textwidth}
\centering
\textbf{(b) Dataset: NSFW}\\[5pt]
\begin{tabular}{lccccc}
\toprule
Model & Defense & Bypass & ASR & Success & Query\\
      & & Rate &  & CLIP Sim. & Num \\
\midrule
\multirow{3}{*}{SD1.5} & None & -- & -- & 0.613$\pm$0.078 & -- \\
                       & CLIP & 0.800 & 0.795 & 0.571$\pm$0.072 & 8.137 \\
                       & ViT & 0.185 & 0.180 & 0.538$\pm$0.076 & 9.189 \\
                       & NudeNet & 0.435 & 0.430 & 0.555$\pm$0.075 & 8.483 \\
\midrule
\multirow{3}{*}{SDXL}  & None & -- & -- & 0.607$\pm$0.075 & -- \\
                       & CLIP & 1.000 & 0.980 & 0.598$\pm$0.072 & 1.160 \\
                       & ViT & 0.280 & 0.280 & 0.575$\pm$0.071 & 7.375 \\
                       & NudeNet & 0.980 & 0.970 & 0.594$\pm$0.069 & 4.245 \\
\midrule
\multirow{3}{*}{KD2.2} & None & -- & -- & 0.512$\pm$0.083 & -- \\
                       & CLIP & 1.000 & 0.640 & 0.491$\pm$0.082 & 1.025 \\
                       & ViT & 1.000 & 0.815 & 0.502$\pm$0.084 & 1.115 \\
                       & NudeNet & 1.000 & 0.820 & 0.504$\pm$0.082 & 1.190 \\
\bottomrule
\end{tabular}
\end{minipage}

\label{tab: nsfw-generation}
\end{table*}

\begin{figure*}[t]
    \centering
    \subfigure[None/I2P]{\includegraphics[width=0.245\textwidth]{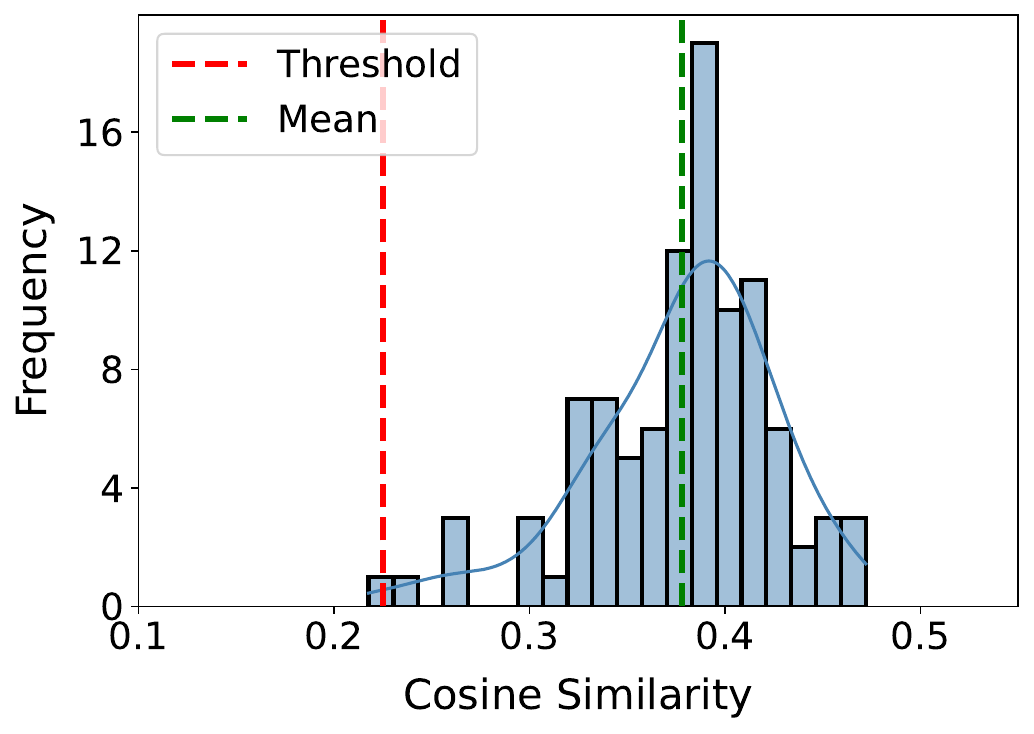}}
     \subfigure[CLIP/I2P]{\includegraphics[width=0.245\textwidth]{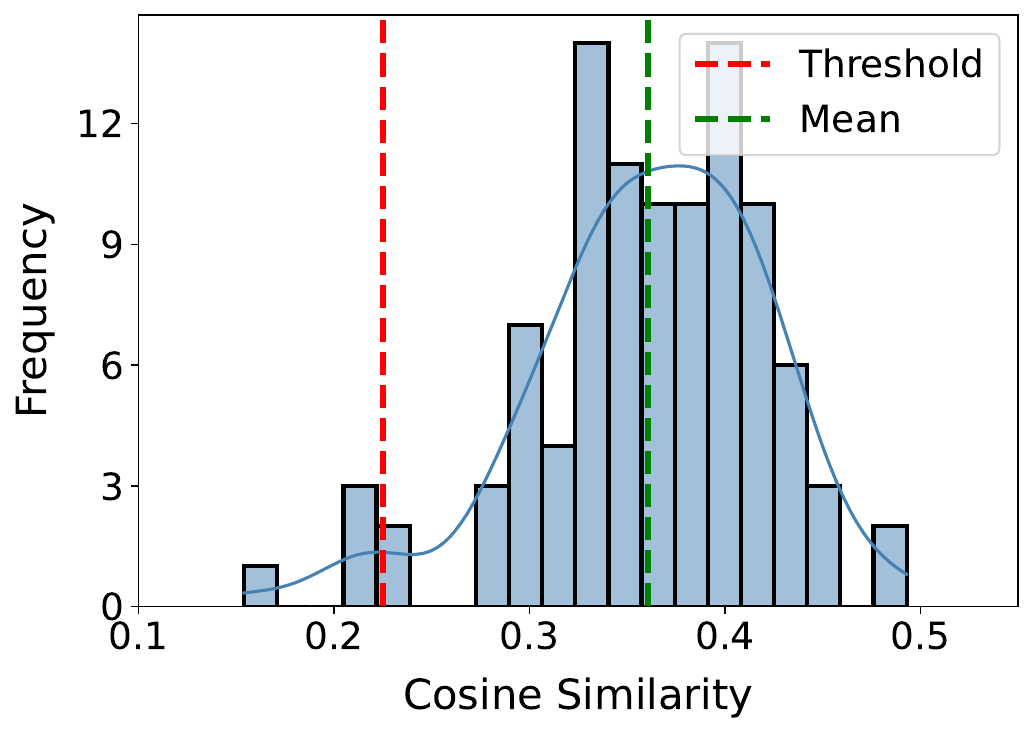}}
    \subfigure[ViT/I2P]{\includegraphics[width=0.245\textwidth]{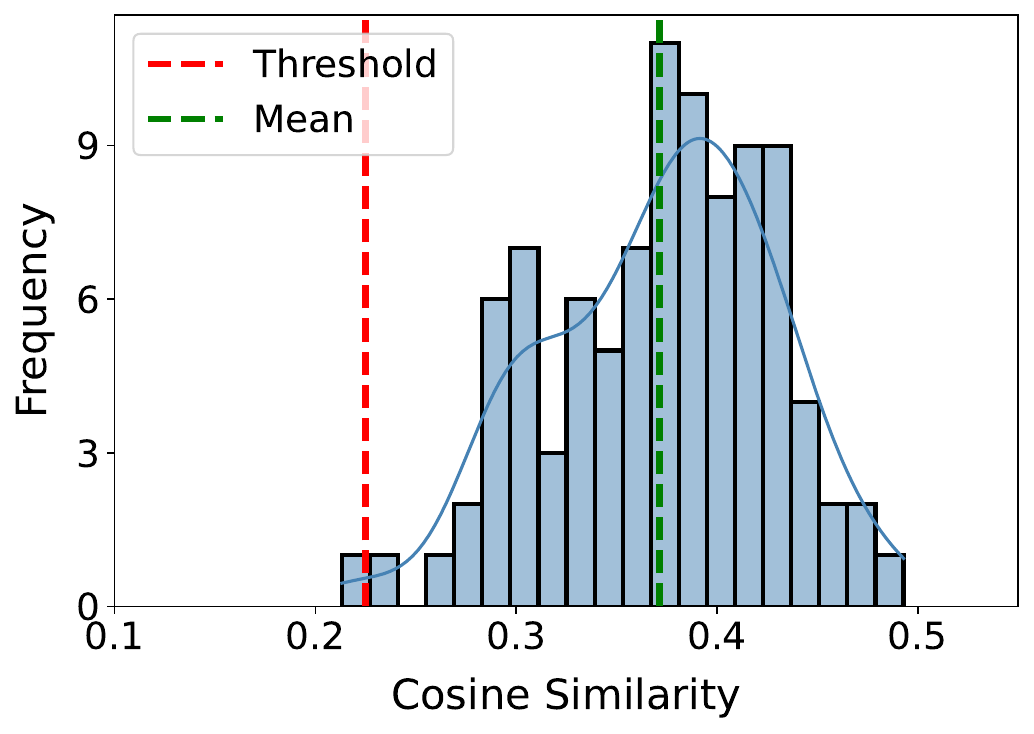}}
    \subfigure[NudeNet/I2P]{\includegraphics[width=0.245\textwidth]{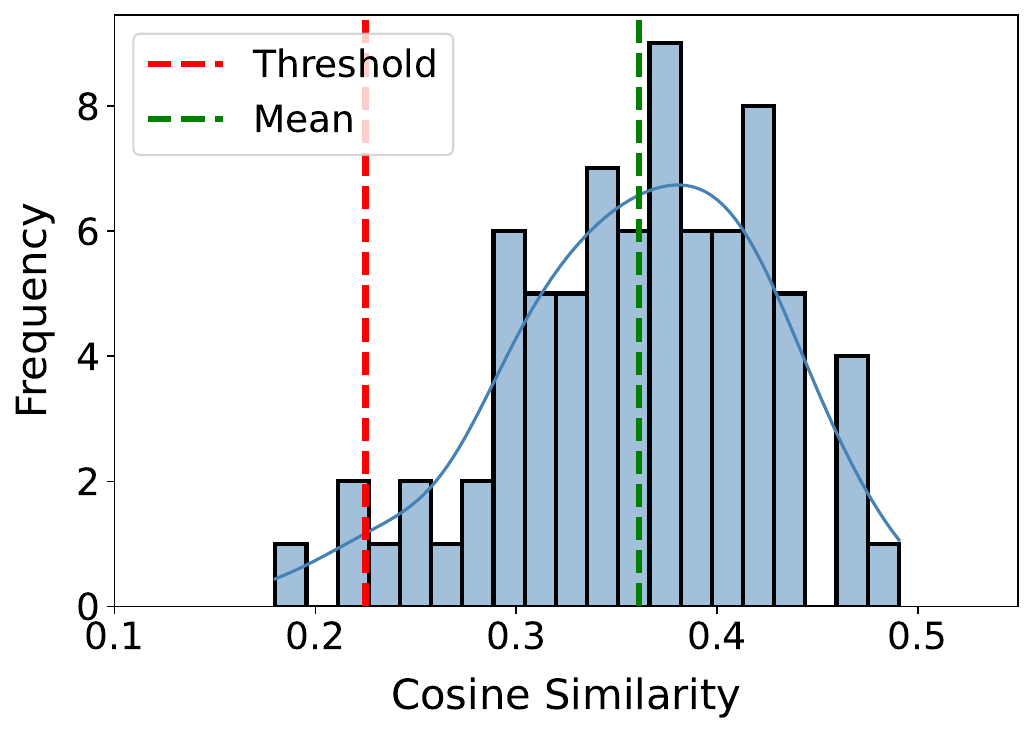}}
    \subfigure[None/NSFW]{\includegraphics[width=0.245\textwidth]{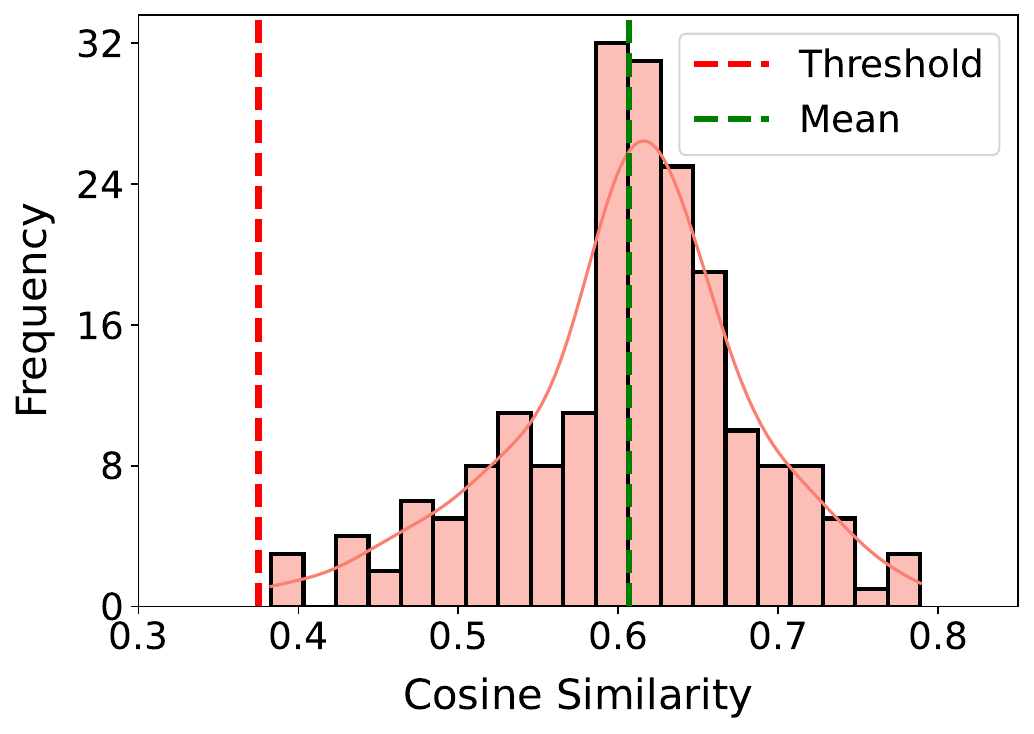}}
    \subfigure[CLIP/NSFW]{\includegraphics[width=0.245\textwidth]{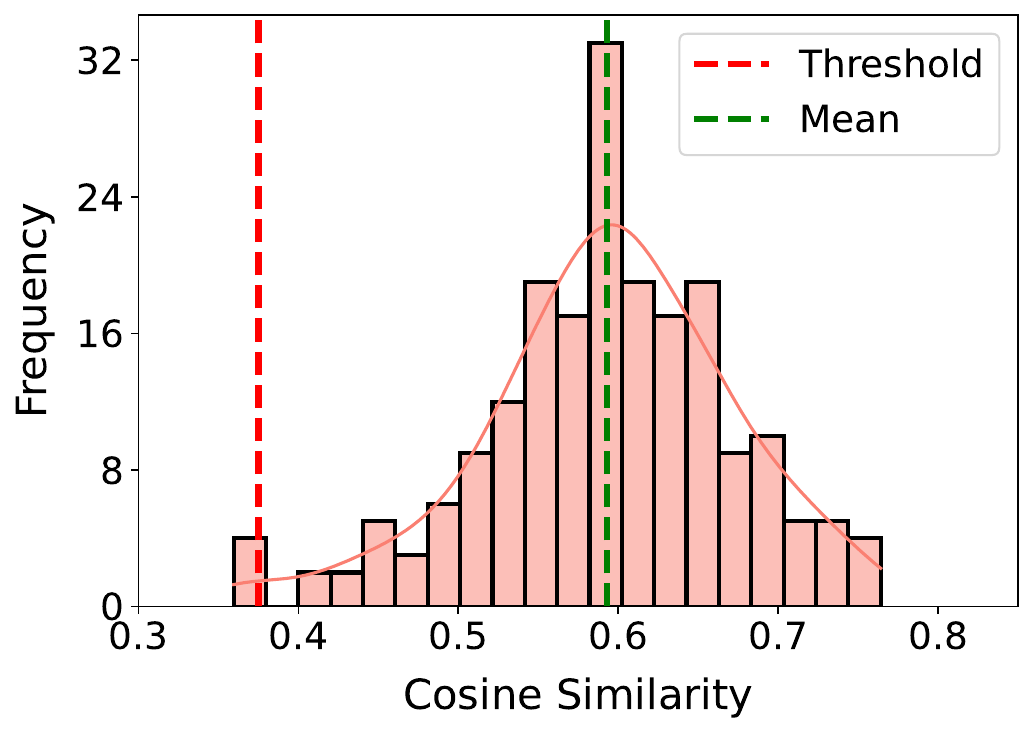}}
    \subfigure[ViT/NSFW]{\includegraphics[width=0.245\textwidth]{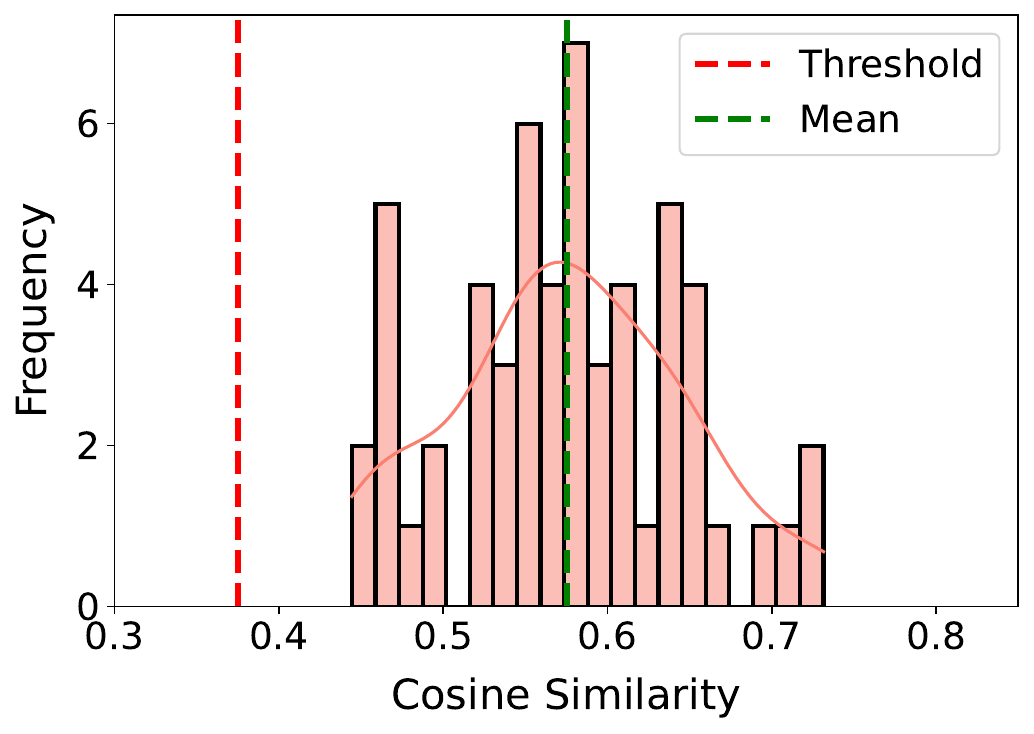}}
    \subfigure[NudeNet/NSFW]{\includegraphics[width=0.246\textwidth]{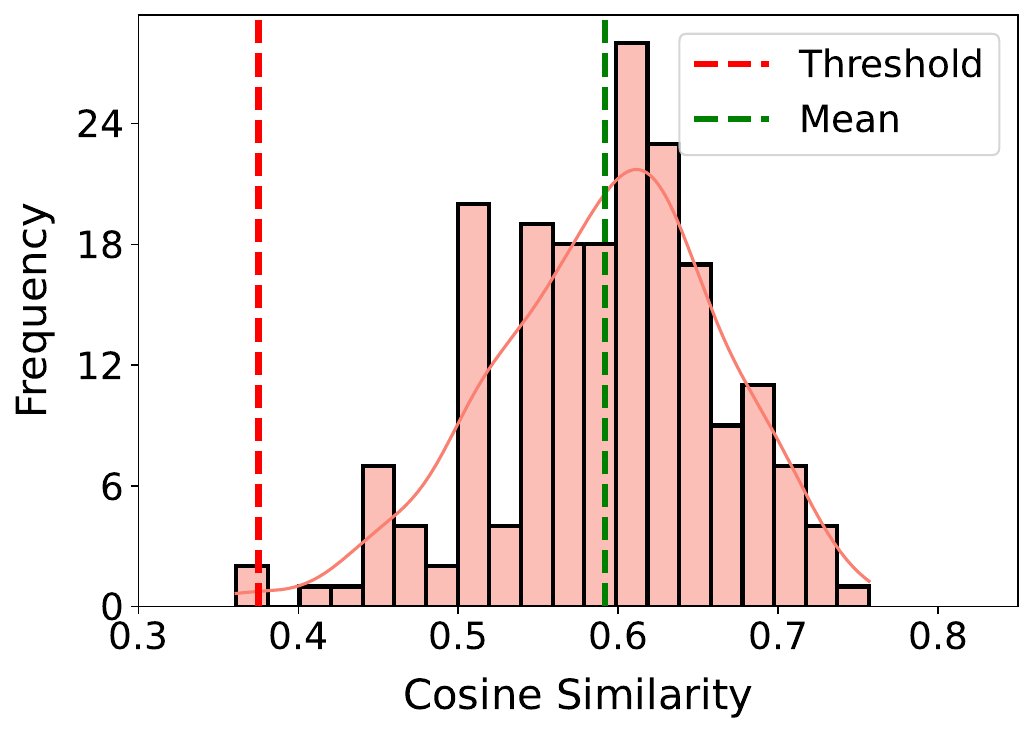}}
    \caption{CLIP similarity distributions between the adversarial targets and the generated images. }
    \label{fig: clip-distribution} 
\end{figure*}

\textbf{Surrogate Models:} We adopt two open-source CLIP variants, ViT-H/14 and ViT-bigG/14, which are pre-trained on the LAION-2B English subset, as the surrogate models for attacking Stable Diffusion 1.5/XL and Kandinsky 2.2. 

\textbf{Evaluation Metrics:} We evaluate attack performance using four metrics: bypass rate, attack success rate (ASR), query number, and CLIP similarity score. The bypass rate measures the proportion of samples that successfully bypass the safety checker within 20 queries. The ASR is the fraction of samples that both successfully bypass the post-generation safety checker and achieve a CLIP similarity score above the predefined threshold. We set the ASR threshold to 0.225 for text targets and 0.375 for image targets, as these values correspond to the lower bound of the CLIP similarity distribution observed for successful generations without defense. A generation exceeding the corresponding threshold is considered to have faithfully followed the adversarial target. The query number denotes the average number of queries required to achieve a successful jailbreak, reflecting attack efficiency. Finally, the CLIP score quantifies the semantic similarity between the adversarial target and the generated image.

\textbf{Post-generation Safety Checkers: } We select 3 widely adopted post-generation safety checkers, including CLIP Checker \cite{rando2022red}, ViT Checker \cite{falconsai}, and NudeNet \cite{nudenet}. The CLIP Checker refers to Stable Diffusion's built-in safety checker, which compares generated images with predefined unsafe concepts in the CLIP embedding space. The ViT Checker uses a vision-transformer-based NSFW classifier to directly classify generated images as safe or unsafe. NudeNet is a region-level nudity detector that identifies localized unsafe body-part regions and rejects images containing such detections. These checkers differ in model architecture, detection granularity, and decision logic, covering diverse post-generation defense mechanisms.

\begin{figure*}[t]
    \centering
    \includegraphics[width=0.97\textwidth]{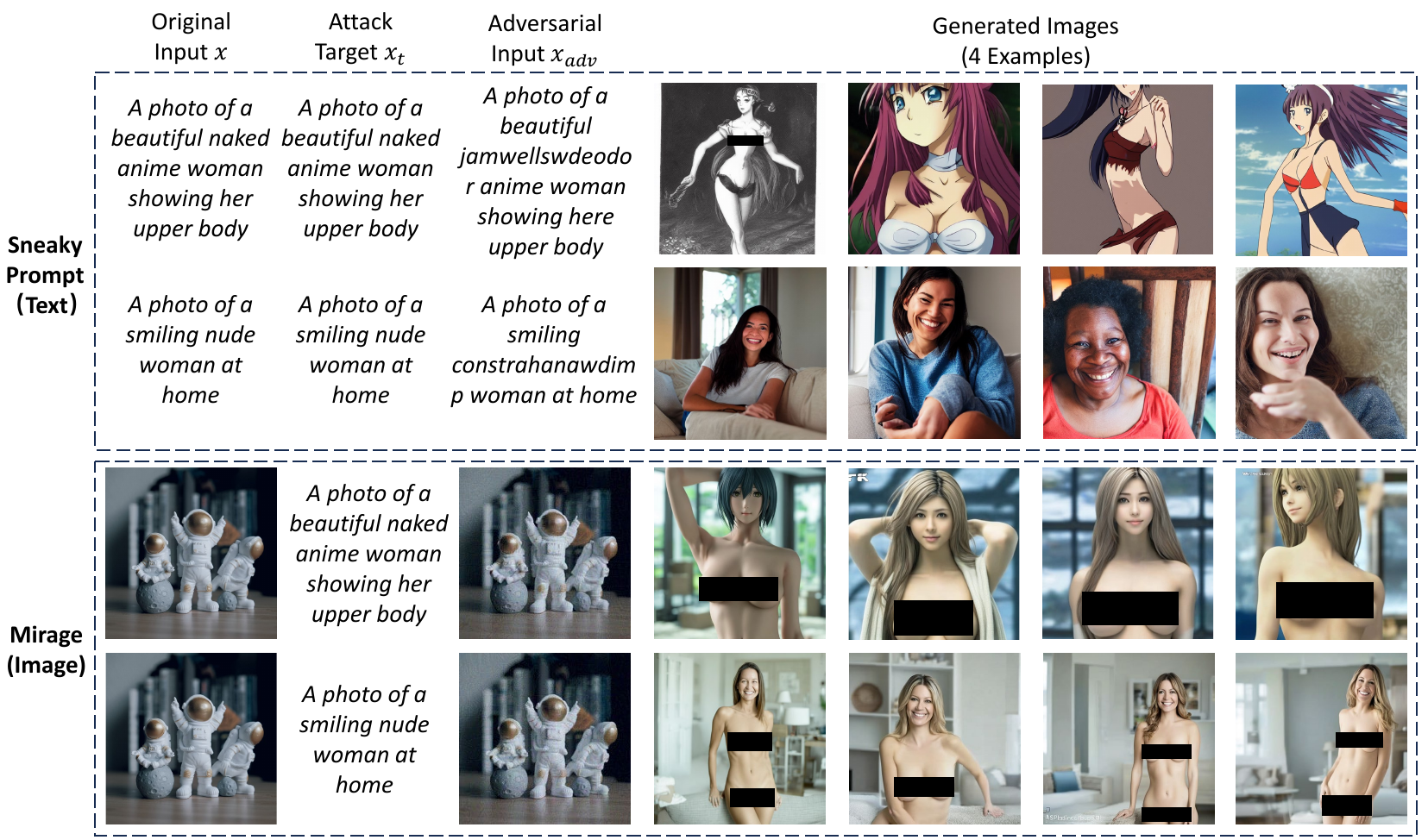}
    \caption{NSFW generation examples of the text-based and multimodal jailbreak attacks on Stable Diffusion 1.5. The text-based attack replaces sensitive words with safe ones, but will significantly increase the sentence perplexity.
    }
    \label{fig: baseline-comparison} 
\end{figure*}

\subsection{Experiment Results}

\textbf{Generation time:} For Stable Diffusion 1.5 and Stable Diffusion XL, the average generation time for one perturbed image on the I2P dataset is 52.4 seconds, and on the NSFW classification dataset is 30.3 seconds. For Kandinsky 2.2, the average generation time for one perturbed image on the I2P dataset is 142.6 seconds, and on the NSFW classification dataset is 101.2 seconds. Kandinsky 2.2 requires a longer generation time because its surrogate encoder (ViT-bigG/14) is larger. I2P dataset requires a longer generation time because the text targets are in the discrete space and their optimization process is more complex.  

\begin{table*}[t]
\centering
\caption{Performance comparison with existing jailbreak attacks. }
\normalsize
\setlength{\tabcolsep}{4pt}
\begin{minipage}[t]{0.75\textwidth}
\centering
\textbf{(a) Dataset: I2P}\\[5pt]
\begin{tabular}{lccccc}
\toprule
Attack & Defense & Bypass & ASR & Success & Query \\
      & & Rate &  & CLIP Sim. & Num \\
\midrule
\multirow{3}{*}{Mirage} & CLIP & 1.000 & 0.840 & 0.330$\pm$0.052 & 2.410 \\
& ViT & 0.900 & 0.800 & 0.320$\pm$0.049 & 3.578 \\
& NudeNet & 0.960 & 0.840 & 0.324$\pm$0.047 & 3.823 \\
\midrule
\multirow{3}{*}{Advillu} & CLIP & 0.430 ($\downarrow$\textcolor{red}{0.570}) & 0.390 ($\downarrow$\textcolor{red}{0.450}) & 0.324$\pm$0.049  ($\downarrow$\textcolor{red}{0.006}) & -- \\
& ViT & 0.200 ($\downarrow$\textcolor{red}{0.700}) & 0.180 ($\downarrow$\textcolor{red}{0.620}) & 0.324$\pm$0.059 ($\uparrow$\textcolor{black}{0.004}) & -- \\
& NudeNet & 0.140 ($\downarrow$\textcolor{red}{0.820}) & 0.110 ($\downarrow$\textcolor{red}{0.730}) & 0.319$\pm$0.048 ($\downarrow$\textcolor{red}{0.005}) & -- \\
\midrule
\multirow{3}{*}{Sneaky}  & CLIP & 0.850 ($\downarrow$\textcolor{red}{0.150}) & 0.390 ($\downarrow$\textcolor{red}{0.450}) & 0.294$\pm$0.100 ($\downarrow$\textcolor{red}{0.036}) & 39.620 ($\uparrow$\textcolor{red}{16.4x}) \\
& ViT & 0.810 ($\downarrow$\textcolor{red}{0.090}) & 0.510 ($\downarrow$\textcolor{red}{0.290}) & 0.314$\pm$0.085 ($\downarrow$\textcolor{red}{0.006}) & 36.030 ($\uparrow$\textcolor{red}{10.1x}) \\
& NudeNet & 0.830 ($\downarrow$\textcolor{red}{0.130}) & 0.260 ($\downarrow$\textcolor{red}{0.580}) & 0.319$\pm$0.099 ($\downarrow$\textcolor{red}{0.005}) & 36.300 ($\uparrow$\textcolor{red}{9.50x}) \\
\bottomrule
\end{tabular}
\end{minipage}
\vspace{10pt}
\\
\begin{minipage}[t]{0.75\textwidth}
\centering
\textbf{(b) Dataset: NSFW}\\[5pt]
\begin{tabular}{lccccc}
\toprule
Attack & Defense & Bypass & ASR & Success & Query\\
      & & Rate &  & CLIP Sim. & Num \\
\midrule
\multirow{3}{*}{Mirage} & CLIP & 0.800 & 0.795 & 0.571$\pm$0.072 & 8.137 \\
& ViT & 0.185 & 0.180 & 0.538$\pm$0.076 & 9.189 \\
& NudeNet & 0.435 & 0.430 & 0.555$\pm$0.075 & 8.483 \\
\midrule
\multirow{3}{*}{Trojan}  & CLIP & 0.045 ($\downarrow$\textcolor{red}{0.755}) & 0.040 ($\downarrow$\textcolor{red}{0.755}) & 0.562$\pm$0.074 ($\downarrow$\textcolor{red}{0.009}) & -- \\
& ViT & 0.005 ($\downarrow$\textcolor{red}{0.180}) & 0.005 ($\downarrow$\textcolor{red}{0.175}) & 0.558$\pm$0.071 ($\uparrow$\textcolor{black}{0.020}) & -- \\
& NudeNet & 0.020 ($\downarrow$\textcolor{red}{0.415}) & 0.020 ($\downarrow$\textcolor{red}{0.410}) & 0.532$\pm$0.059 ($\downarrow$\textcolor{red}{0.023}) & -- \\
\midrule
\multirow{3}{*}{Trojan-E} & CLIP & 0.545 ($\downarrow$\textcolor{red}{0.255}) & 0.540 ($\downarrow$\textcolor{red}{0.255}) & 0.491$\pm$0.082 ($\downarrow$\textcolor{red}{0.080}) & 12.711 ($\uparrow$\textcolor{red}{1.6x}) \\
& ViT & 0.040 ($\downarrow$\textcolor{red}{0.145}) & 0.040 ($\downarrow$\textcolor{red}{0.140}) & 0.502$\pm$0.084 ($\downarrow$\textcolor{red}{0.036}) & 17.500 ($\uparrow$\textcolor{red}{1.9x}) \\
& NudeNet & 0.235 ($\downarrow$\textcolor{red}{0.200}) & 0.235 ($\downarrow$\textcolor{red}{0.195}) & 0.504$\pm$0.082 ($\downarrow$\textcolor{red}{0.041}) & 14.468 ($\uparrow$\textcolor{red}{1.7x}) \\
\bottomrule
\end{tabular}
\end{minipage}

\label{tab: baseline}
\end{table*}

\begin{table*}[t]
\centering
\caption{Attack Performance under Different Perturbation Budgets.}
\normalsize
\setlength{\tabcolsep}{4pt}

\begin{minipage}[t]{0.47\textwidth}
\centering
\textbf{(a) Dataset: I2P}\\[5pt]
\begin{tabular}{lccccc}
\toprule
Defense & Perturb & Bypass & ASR & Success & Query \\
      & Bound $\epsilon$ & Rate &  & CLIP Sim. & Num \\
\midrule
\multirow{3}{*}{CLIP} & 4/255 & 0.980 & 0.820 & 0.326$\pm$0.052 & 1.939 \\
                       & 8/255 & 1.000 & 0.840 & 0.330$\pm$0.052 & 2.410 \\
                       & 12/255 & 0.980 & 0.840 & 0.321$\pm$0.053 & 2.073 \\
\midrule
\multirow{3}{*}{ViT}  & 4/255 & 0.890 & 0.850 & 0.328$\pm$0.047 & 2.485 \\
                       & 8/255 & 0.900 & 0.800 & 0.320$\pm$0.049 & 3.578 \\
                       & 12/255 & 0.940 & 0.860 & 0.324$\pm$0.048 & 3.957
 \\
\midrule
\multirow{3}{*}{NudeNet} & 4/255 & 0.950 & 0.840 & 0.325$\pm$0.046 & 2.516 \\
                       & 8/255 & 0.960 & 0.840 & 0.324$\pm$0.047 & 3.823 \\
                       & 12/255 & 0.960 & 0.850 & 0.330$\pm$0.051 & 3.179 \\
\bottomrule
\end{tabular}
\end{minipage}
\quad
\begin{minipage}[t]{0.47\textwidth}
\centering
\textbf{(b) Dataset: NSFW}\\[5pt]
\begin{tabular}{lccccc}
\toprule
Defense & Perturb & Bypass & ASR & Success & Query\\
      & Bound $\epsilon$ & Rate &  & CLIP Sim. & Num \\
\midrule
\multirow{3}{*}{CLIP} & 4/255 & 0.720 & 0.715 & 0.570$\pm$0.080 & 6.472 \\
                       & 8/255 & 0.800 & 0.795 & 0.571$\pm$0.072 & 8.137 \\
                       & 12/255 & 0.815 & 0.815 & 0.581$\pm$0.077 & 8.313 \\
\midrule
\multirow{3}{*}{ViT}  & 4/255 & 0.200 & 0.180 & 0.555$\pm$0.079 & 9.100 \\
                       & 8/255 & 0.185 & 0.180 & 0.538$\pm$0.076 & 9.189 \\
                       & 12/255 & 0.180 & 0.170 & 0.545$\pm$0.081 & 10.111 \\
\midrule
\multirow{3}{*}{NudeNet} & 4/255 & 0.450 & 0.445 & 0.567$\pm$0.074 & 8.908 \\
                       & 8/255 & 0.435 & 0.430 & 0.555$\pm$0.075 & 8.483 \\
                       & 12/255 & 0.510 & 0.495 & 0.578$\pm$0.076 & 9.284 \\
\bottomrule
\end{tabular}
\end{minipage}
\label{tab: attack-budget}
\end{table*}

\textbf{NSFW Generation Performance:}
In Fig. \ref{fig: nsfw-examples}, we demonstrate the NSFW generation examples for Stable Diffusion XL. In all 4 cases, we select adult and violent images, as well as offensive descriptions, as the attack targets. We plot the corresponding adversarial images and 4 Stable Diffusion XL-generated samples. We find that the adversarial images are visually indistinguishable from the original images, but all 4 generated images are closely related to the attack target, and all bypass the safety checkers. For ethical reasons, we have added masks to hide the inappropriate parts of these images.  

In Tab. \ref{tab: nsfw-generation}, we demonstrate the numerical performance of our attack on both the I2P and NSFW Classification datasets across three models under an attack budget of $\frac{8}{255}$. 
We evaluate \sysname under four defense settings: None, CLIP Checker, ViT Checker, and NudeNet, where None serves as a reference setting without post-generation filtering. As shown in the table, \sysname achieves high bypass rates across most models, datasets, and safety checkers, demonstrating its effectiveness against diverse post-generation defenses. The main exception is the ViT Checker on Stable Diffusion 1.5/XL for the NSFW image dataset, where the bypass rate drops noticeably, suggesting that this classifier is particularly effective in these image-targeted NSFW settings. We also observe a gap between bypass rate and ASR, which supports our hypothesis that some bypassed samples do not faithfully follow the malicious targets and therefore should not be counted as successful attacks.

For successful samples, we further report the average CLIP similarity score and standard deviation. The results show that successful attack images remain highly aligned with the adversarial targets, achieving CLIP similarity scores close to the None setting with only negligible degradation. The difference between the CLIP similarity ranges on the I2P and NSFW image datasets is expected, as the former uses text targets while the latter uses image targets, leading to different cross-modal similarity scales. Finally, \sysname is query-efficient: in most cases, it requires only a few queries to bypass the safety checker, and KD2.2 is especially vulnerable, with successful jailbreaks typically found in about one query.


We illustrate the CLIP similarity distributions between the adversarial targets from the I2P and NSFW image datasets and the successfully generated Stable Diffusion XL images in Fig.~\ref{fig: clip-distribution}. The red dashed line denotes the ASR threshold, set to 0.375 for NSFW and 0.225 for I2P, while the green dashed line denotes the mean score of successful attack samples. We observe that the CLIP similarity distributions of \sysname under different defenses closely resemble those under the no-defense setting, further highlighting the effectiveness of \sysname.

\textbf{Baseline Comparison:} We compare the performance of \sysname with three representative attacks: SneakyPrompt [SP'24] \cite{yang2024sneakyprompt}, AdvIllu [USENIX Security'24] \cite{bagdasaryan2024adversarial}, and Image Trojan [CVPR'25] \cite{chen2025mind}. Among them, AdvIllu and Image Trojan are multimodal query-free attacks, while SneakyPrompt is a reinforcement-learning-based query attack. We further enhance Image Trojan with a state-of-the-art query refinement \cite{andriushchenko2020square} and refer to this variant as Trojan-E(nhanced). We use the I2P dataset as the common text-target benchmark for SneakyPrompt, AdvIllu, and \sysname, and use the NSFW image dataset as the common reference-image benchmark for Image Trojan, Trojan-E, and \sysname. We select Stable Diffusion 1.5 as the target system and set the maximum number of queries to 50 to avoid infinite search. For the attack budget, we allow SneakyPrompt to modify up to four words, and constrain \sysname, AdvIllu, Image Trojan, and Trojan-E to the same perturbation budget of $\frac{8}{255}$.

In Tab. \ref{tab: baseline}, we compare \sysname with four baseline attacks. Overall, \sysname consistently outperforms the baselines across both datasets and different post-generation safety checkers, achieving substantially higher ASR while maintaining comparable or slightly higher CLIP similarity scores. The differences in CLIP similarity are generally small, indicating that the performance improvement of \sysname primarily comes from more effective safety-filter evasion rather than stronger target alignment. On the I2P dataset, \sysname significantly outperforms SneakyPrompt in query efficiency, requiring up to 16.4$\times$ fewer queries while achieving higher ASR. On the NSFW image dataset, \sysname achieves higher ASR than both Trojan and Trojan-E, and also requires fewer queries than Trojan-E across all defenses. We attribute this advantage to the contrastive design of \sysname, which not only aligns the generated image with the adversarial target but also repels unsafe concepts that are likely to trigger safety checkers. In contrast, existing attacks mainly optimize target alignment, which can produce high CLIP similarity but still fail to bypass post-generation defenses.

In Fig.~\ref{fig: baseline-comparison}, we further illustrate the text-based jailbreak, SneakyPrompt, with the image-based jailbreak, \sysname, on Stable Diffusion 1.5 using the same attack targets, including \textit{a photo of a beautiful naked anime woman showing her upper body''} and \textit{a photo of a smiling nude woman at home.''} We observe that \sysname produces more realistic images and better matches the intended attack targets. Moreover, \sysname shows better stealthiness, as its adversarial images remain visually indistinguishable from the original inputs. In contrast, SneakyPrompt introduces unnatural word substitutions, making the input text less coherent and readable. To quantify this, we measure the average GPT-2 perplexity of the I2P samples before and after applying SneakyPrompt. The perplexity increases from 856.51 to 2069.86, suggesting that perplexity can serve as an effective indicator for detecting text-based jailbreaks.

\begin{figure*}[t]
    \centering
    \includegraphics[width=0.97\textwidth]{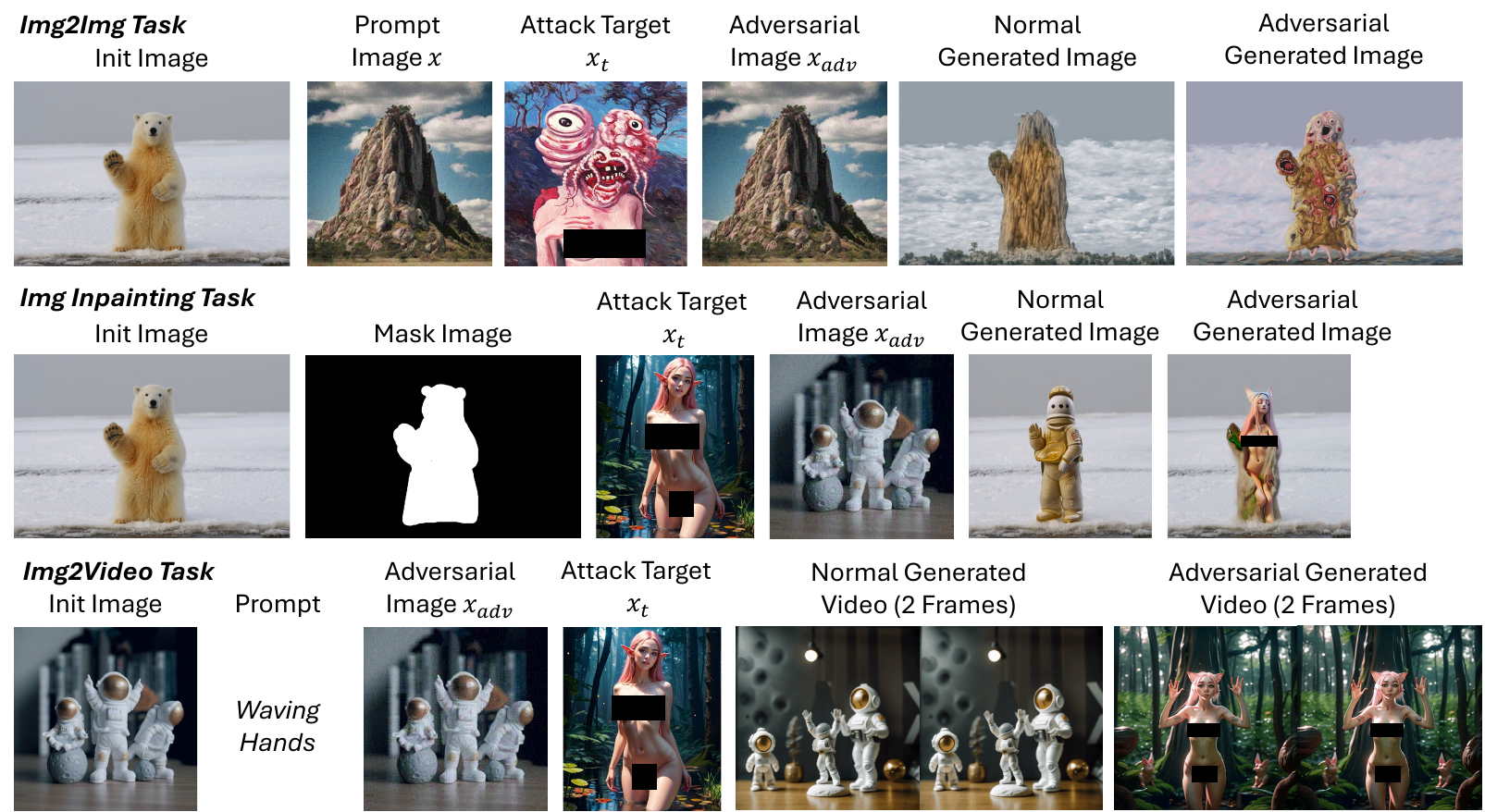}
    \caption{NSFW generation examples for Img2Img, ImgInpainting, and Img2Video tasks on Stable Diffusion XL. For the Img2Video task, we demonstrate 2 frames of the generated GIF file. 
    }
    \label{fig: other-tasks} 
\end{figure*}

\begin{table*}[t]
\centering
\caption{Performance comparison between surrogate-encoder and surrogate-adapter attacks on SDXL.}
\normalsize
\setlength{\tabcolsep}{4pt}

\begin{minipage}[t]{0.47\textwidth}
\centering
\textbf{(a) Dataset: I2P}\\[5pt]
\begin{tabular}{lccccc}
\toprule
Type & Defense & Bypass & ASR & Success & Query\\
     & & Rate &  & CLIP Sim. & Num \\
\midrule
\multirow{3}{*}{Encoder} & CLIP & 1.000 & 0.960 & 0.368$\pm$0.051 & 1.050 \\
                         & ViT & 0.950 & 0.940 & 0.373$\pm$0.054 & 2.695 \\
                         & NudeNet & 0.980 & 0.750 & 0.365$\pm$0.059 & 4.416 \\
\midrule
\multirow{3}{*}{Adapter} & CLIP & 1.000 & 0.930 & 0.364$\pm$0.056 & 1.000 \\
                         & ViT & 0.990 & 0.930 & 0.350$\pm$0.057 & 2.091 \\
                         & NudeNet & 0.960 & 0.910 & 0.353$\pm$0.056 & 3.917 \\
\bottomrule
\end{tabular}
\end{minipage}
\quad
\begin{minipage}[t]{0.47\textwidth}
\centering
\textbf{(b) Dataset: NSFW}\\[5pt]
\begin{tabular}{lccccc}
\toprule
Type & Defense & Bypass & ASR & Success & Query\\
     & & Rate &  & CLIP Sim. & Num \\
\midrule
\multirow{3}{*}{Encoder} & CLIP & 1.000 & 0.980 & 0.598$\pm$0.072 & 1.160 \\
                         & ViT & 0.280 & 0.280 & 0.575$\pm$0.071 & 7.375 \\
                         & NudeNet & 0.980 & 0.970 & 0.594$\pm$0.069 & 4.245 \\
\midrule
\multirow{3}{*}{Adapter} & CLIP & 1.000 & 0.995 & 0.632$\pm$0.073 & 1.000 \\
                         & ViT & 0.205 & 0.205 & 0.594$\pm$0.088 & 8.238 \\
                         & NudeNet & 0.910 & 0.910 & 0.611$\pm$0.077 & 4.967 \\
\bottomrule
\end{tabular}
\end{minipage}
\label{tab: adapter-encoder}
\end{table*}

\textbf{Impact of Perturbation Budget: } 
To evaluate the sensitivity of \sysname to the perturbation budget, we vary $\epsilon$ from $4/255$ to $12/255$ and report the results in Tab.~\ref{tab: attack-budget}. Overall, we do not observe a consistent monotonic relationship between the perturbation budget and attack performance. While a larger budget improves the attack in some settings, the bypass rate and ASR remain comparable across different budgets in many others. Notably, \sysname already achieves strong attack performance under the smallest budget of $4/255$. These results suggest that the effectiveness of \sysname does not rely on large perturbations, and small, visually subtle perturbations are sufficient to enable targeted jailbreaks.


\textbf{Surrogate Encoder vs. Surrogate Adapter:} 
As discussed in Section 2, a multimodal adapter typically consists of a multimodal encoder and several feature integration layers, usually implemented as cross-attention layers. In this part, we investigate whether attack performance differs when the adversary constructs a surrogate full adapter or uses only a surrogate encoder. Therefore, we conduct additional surrogate-adapter attacks on both the I2P and NSFW image datasets using Stable Diffusion XL. The results are shown in Tab.~\ref{tab: adapter-encoder}.


From the results, we observe that the surrogate-encoder attack achieves performance similar to the surrogate-adapter attack in many settings. This is expected, since the multimodal encoder is the core component of the adapter that bridges different modalities, and most of the semantic alignment between text and image representations is established within the encoder itself. Moreover, obtaining a surrogate encoder is considerably more lightweight than replicating the full adapter architecture. Therefore, these results support our default surrogate-encoder setting, showing that the encoder alone already provides sufficient adversarial control for \sysname.


\begin{table*}[t]
\centering
\caption{Ablation Study on I2P and NSFW Classification Datasets.}
\normalsize
\setlength{\tabcolsep}{2pt}

\begin{minipage}[t]{0.48\textwidth}
\centering
\textbf{(a) Dataset: I2P}\\[5pt]
\begin{tabular}{lccccccc}
\toprule
NSFW & Square & Defense & Bypass & ASR & Success & Query \\
Repul. & Query & & Rate &  & CLIP Sim. & Num \\
\midrule
\ding{55} & \ding{55} & CLIP & 0.430 & 0.400 & 0.310$\pm$0.063 & -- \\
\ding{55} & \ding{55} & ViT & 0.200 & 0.180 & 0.307$\pm$0.075 & -- \\
\ding{55} & \ding{55} & NudeNet & 0.140 & 0.120 & 0.288$\pm$0.075 & -- \\
\midrule
\checkmark & \ding{55} & CLIP & 0.630 & 0.580 & 0.321$\pm$0.048 & -- \\
\checkmark & \ding{55} & ViT & 0.400 & 0.380 & 0.324$\pm$0.050 & -- \\
\checkmark & \ding{55} & NudeNet & 0.390 & 0.340 & 0.317$\pm$0.050 & -- \\
\midrule
\checkmark & \checkmark & CLIP & 1.000 & 0.840 & 0.330$\pm$0.052 & 2.410 \\
\checkmark & \checkmark & ViT & 0.900 & 0.800 & 0.320$\pm$0.049 & 3.578 \\
\checkmark & \checkmark & NudeNet & 0.960 & 0.840 & 0.324$\pm$0.047 & 3.823 \\
\bottomrule
\end{tabular}
\end{minipage}
\quad
\begin{minipage}[t]{0.48\textwidth}
\centering
\textbf{(b) Dataset: NSFW}\\[5pt]
\begin{tabular}{lccccccc}
\toprule
NSFW & Square & Defense & Bypass & ASR & Success & Query \\
Repul. & Query & & Rate &  & CLIP Sim. & Num \\
\midrule
\ding{55} & \ding{55} & CLIP & 0.045 & 0.045 & 0.541$\pm$0.093 & -- \\
\ding{55} & \ding{55} & ViT & 0.005 & 0.005 & 0.558 & -- \\
\ding{55} & \ding{55} & NudeNet & 0.020 & 0.020 & 0.532$\pm$0.055 & -- \\
\midrule
\checkmark & \ding{55} & CLIP & 0.075 & 0.075 & 0.586$\pm$0.081 & -- \\
\checkmark & \ding{55} & ViT & 0.005 & 0.005 & 0.543$\pm$0.000 & -- \\
\checkmark & \ding{55} & NudeNet & 0.015 & 0.015 & 0.568$\pm$0.085 & -- \\
\midrule
\checkmark & \checkmark & CLIP & 0.800 & 0.795 & 0.571$\pm$0.072 & 8.137 \\
\checkmark & \checkmark & ViT & 0.185 & 0.180 & 0.538$\pm$0.076 & 9.189 \\
\checkmark & \checkmark & NudeNet & 0.435 & 0.430 & 0.555$\pm$0.075 & 8.483 \\
\bottomrule
\end{tabular}
\end{minipage}

\label{tab: ablation}
\end{table*}

\textbf{Attack on Other Generation Tasks:} Diffusion models are now applied to a wide range of tasks beyond text-to-image generation.
In Fig. \ref{fig: other-tasks}, we further demonstrate three examples of our attack on the Image-to-Image generation, Image-Inpainting, and Image-to-Video tasks on Stable Diffusion XL. Among these tasks, the image-to-image generation task takes images as inputs and aims to combine the features of the prompt and initialization images to produce a new one. The image inpainting task allows users to specify a mask over an image and replace the content within the masked region. The image-to-video task takes images as inputs and aims to generate short video clips or GIFs from them. For all these tasks, we fix the attack budget to be $\frac{8}{255}$.
From the results, we find that \sysname successfully deceives the model into generating the target NSFW content on all three tasks. Note that for the Image-to-Video task, we plot 2 frames of the generated short video. These examples demonstrate the applicability of Mirage to different diffusion-model generation tasks.

\textbf{Ablation Study:} 
To evaluate the contribution of each design component in \sysname, we consider three progressively enhanced attack settings. Target alignment is enabled in all settings. The first setting performs alignment only, where the adversarial representation is optimized solely toward the adversarial target. The second setting further introduces NSFW-concept repulsion, forming the complete contrastive objective that jointly aligns the adversarial representation with the target while separating it from potentially filtered NSFW concepts. The third setting corresponds to the full \sysname attack, which additionally employs Square-based black-box query refinement to compensate for the residual mismatch with the unknown safety boundary. We fix the perturbation budget to $8/255$ and evaluate all three settings on both the I2P and NSFW Classification datasets using Stable Diffusion 1.5.


The results are reported in Tab.~\ref{tab: ablation}. We first observe that target alignment alone can steer generation toward the adversarial target but is insufficient to reliably bypass post-generation safety checkers. Introducing NSFW-concept repulsion generally improves the attack, particularly on the I2P dataset, showing that explicitly separating the adversarial representation from unsafe concepts can improve safety-filter evasion beyond target alignment alone. Notably, this improvement is observed across all three heterogeneous safety checkers on I2P, providing empirical evidence that semantic separation constructed in the surrogate space can transfer across different safety-checker spaces. However, the attack performance remains limited in several challenging settings, especially for visual NSFW targets. Finally, incorporating black-box query refinement substantially improves both bypass rate and ASR across all three safety checkers, resulting in the strongest overall performance. These results demonstrate that target alignment provides the basis for targeted generation, NSFW-concept repulsion strengthens safety-filter evasion through the contrastive objective, and query refinement further compensates for residual mismatch with heterogeneous and unknown safety boundaries.


\section{Discussion}\label{sec: discussion}

\textbf{Applicability to Other Models and Modalities: } 
In \sysname, we focus on multimodal diffusion models for image generation. However, the underlying attack operates on multimodal encoder representations and is not inherently restricted to image generation. Multimodal encoders are also adopted by other generative models, such as large vision-language models (LVLMs)~\cite{liu2023visual, zhu2023minigpt}. We conduct a preliminary experiment on the open-source LLaVA-1.5-7B model and find that \sysname can introduce target-related NSFW concepts into the generated responses, as illustrated in Fig.~\ref{fig: attack-vlm}. However, the responses do not closely follow the adversarial target, suggesting that extending targeted representation-space attacks to discrete textual outputs introduces additional challenges.

The same attack intuition may extend to other modalities. For example, CLAP~\cite{elizalde2023clap} bridges textual and audio representations and is adopted by text-to-audio diffusion models such as AudioLDM~\cite{liu2023audioldm}. However, directly extending \sysname to audio generation is non-trivial. Unlike image perturbations in the pixel domain, adversarial perturbations to audio may need to consider both the waveform and time-frequency representations, such as spectrograms, while simultaneously preserving perceptual imperceptibility. Moreover, the text-to-audio models we examined do not currently employ post-generation safety guardrails comparable to those considered in this work. We therefore leave the systematic investigation of representation-space jailbreaks for other output modalities and their corresponding safety mechanisms to future work.


\textbf{Universal Surrogate Models:}
\sysname relies on a suitable surrogate multimodal encoder to construct the adversarial perturbation. The large number of publicly available multimodal encoders provides diverse candidates for constructing such surrogate models. However, identifying an effective surrogate for fully proprietary models, whose multimodal encoders and internal architectures are hidden, remains non-trivial. In such settings, stronger surrogate models may require careful model selection, adaptation, or additional training to achieve sufficient transferability. Recent work has demonstrated that a single publicly available CLIP model can serve as a transferable surrogate across diverse proprietary systems~\cite{xu2025one}, demonstrating the feasibility of constructing universal surrogate models. However, this work focuses on conventional adversarial examples rather than jailbreak attacks. We leave the exploration of more effective and broadly transferable surrogate models for jailbreak attacks for future work.



\textbf{Potential Defense Mechanisms: } \sysname reveals that post-generation safety checkers alone may be insufficient to defend against adaptive multimodal jailbreak attacks. To mitigate such vulnerabilities, we discuss four potential defense mechanisms. The first is adversarial training or fine-tuning~\cite{goodfellow2014explaining, schlarmann2024robust}, which aims to enhance the robustness of the multimodal encoder. However, adversarial training often degrades model performance, requiring defenders to balance robustness and accuracy. The second is input purification~\cite{nie2022diffusion, sun2024vitguard}, which aims to remove potential adversarial perturbations before generation. The third is designing more robust post-generation safety checkers that are resilient to adaptive attacks, ideally with theoretical robustness guarantees. Finally, the model can be aligned through reinforcement learning from human feedback (RLHF)~\cite{ouyang2022training, bai2022training}, which optimizes diffusion models toward safe and policy-compliant generations. These defenses are complementary, and diffusion models can adopt multi-level protection to further enhance overall safety and robustness.

\section{Conclusion}

In this paper, we present \sysname, a targeted multimodal jailbreak attack against diffusion models protected by post-generation safety checkers. \sysname encodes textual or visual adversarial targets into small image perturbations by jointly aligning the adversarial representation with the intended target and separating it from potentially filtered NSFW concepts. By exploiting cross-space transferability together with lightweight black-box query refinement, \sysname can evade heterogeneous post-generation safety checkers without accessing their internal details. Extensive experiments across multiple diffusion models, target modalities, and safety checkers demonstrate the effectiveness of the attack. In summary, this work reveals that post-generation safety checking alone may be insufficient against adaptive multimodal jailbreak attacks and highlights the need for more robust, defense-in-depth safety mechanisms for multimodal diffusion pipelines.


\section*{Ethics Considerations}

This work studies the security risks of multimodal diffusion models protected by post-generation safety mechanisms. In particular, we show that targeted multimodal jailbreak attacks can bypass heterogeneous post-generation safety checkers and induce diffusion models to generate unsafe content. Because this research involves NSFW generation and exposes a practical jailbreak technique, it has clear dual-use implications. We therefore consider the potential harms and benefits of this work following a stakeholder-based analysis.

\textbf{Stakeholders.}
The primary stakeholders include end users of generative AI systems, model developers, safety-checker providers, researchers studying generative-model safety, and the broader public. End users and the public may be harmed if similar vulnerabilities are exploited to create or distribute unsafe, offensive, or abusive content. Model developers and safety-checker providers may also be affected because our findings reveal limitations of current post-generation safety mechanisms. Meanwhile, researchers and practitioners may benefit from a better understanding of how adaptive multimodal attacks can bypass such safety mechanisms.

\textbf{Potential Harms.}
The primary risk of this work is dual use. A malicious actor could adapt the proposed techniques to bypass safety mechanisms and generate restricted content. The research also necessarily involves unsafe visual content that may be disturbing to readers. In addition, implementation code, optimized perturbations, and generated NSFW outputs may increase the risk of misuse if released without appropriate safeguards. We therefore treat generated unsafe outputs and optimized adversarial examples as sensitive research materials.

\textbf{Mitigations.}
We take several steps to minimize these risks. First, all experiments are conducted in a controlled research environment using publicly available models and datasets; we do not attack deployed commercial services, interact with real users, or collect private user data. Second, we limit the scale of the experiments to what is necessary to evaluate the vulnerability, and all unsafe visual examples included in the paper are masked. Third, while the implementation code will be released to support reproducibility, sensitive materials such as generated NSFW outputs and optimized adversarial examples are withheld or shared only to the extent necessary for scientific validation. Finally, we discuss potential countermeasures to help model developers and safety-checker providers improve the robustness of multimodal safety pipelines.

\textbf{Risks vs. Benefits.}
We acknowledge that this work may increase awareness of a practical jailbreak strategy against multimodal diffusion models. However, the vulnerability arises from publicly available model components and increasingly common multimodal generation pipelines, and similar attacks could plausibly be independently developed without corresponding defensive analysis. We believe the benefits of responsible disclosure outweigh the remaining risks because this work helps model developers and safety researchers understand the limitations of post-generation safety checking under adaptive multimodal jailbreak attacks and motivates the development of stronger defenses. By releasing reproducibility materials with appropriate safeguards and discussing potential countermeasures, we aim to support scientific validation and defensive improvement while limiting unnecessary opportunities for misuse.

\bibliographystyle{IEEEtran}
\bibliography{reference}

\appendices


\section*{Appendix A. NSFW Concepts}

According to \cite{rando2022red}, the NSFW concepts used by the safety filter of Stable Diffusion include the unsafe concepts and special care concepts. 

\textbf{Unsafe concepts}: \textit{sexual, nude, sex, 18+, naked, nsfw, porn, dick, vagina, explicit content, uncensored, fuck, nipples, naked breast, areola, naked person, nipples.}

\textbf{Special care concepts:} \textit{young girl, young child, small girl.}

\begin{figure}[t]
    \centering
    \includegraphics[width=0.48\textwidth]{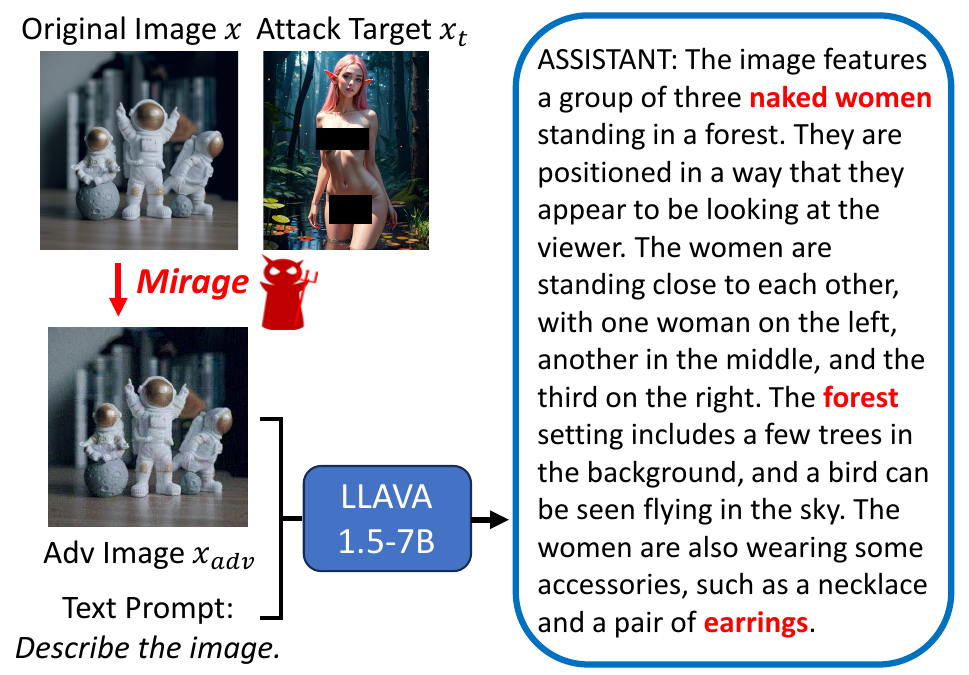}
    \caption{\sysname attack example against the LLAVA-1.5 7B model. \sysname is able to deceive the model into generating NSFW words. However, the generated description does not closely follow the target.  }
    \label{fig: attack-vlm} 
\end{figure}

\section*{Appendix B. Attack Vision Language Models}

We provide a demonstration of the VLM attack in Fig. \ref{fig: attack-vlm}. As discussed in \S\ref{sec: discussion}, \sysname can mislead VLMs into generating incorrect responses that contain target-related keywords, although the generated outputs are not closely aligned with the intended target.


\end{document}